\documentclass[11pt]{article}
\usepackage{graphicx}
\usepackage{subfig}
\usepackage{amsfonts}
\usepackage{amsthm}
\usepackage{amsmath}
\usepackage{amssymb}
\usepackage{dsfont}
\usepackage{enumerate}
\usepackage{natbib}
\usepackage{hyperref}
\usepackage{float}
\usepackage{booktabs}
\usepackage{multirow}
\usepackage{hhline}
\usepackage{framed}
\usepackage{longtable}
\usepackage{mathtools}
\usepackage{rotating}
\usepackage{ragged2e}

\usepackage{tikz}
\usetikzlibrary{calc,arrows,positioning,automata,fit,shapes.geometric,backgrounds,patterns}
\tikzset{
    every node/.style={font=\sffamily\small},
    main node/.style={thick,circle,draw,font=\sffamily\Large}
}

\newtheorem{proposition}{Proposition}
\newtheorem*{proposition*}{Proposition}

\newtheorem{assumption}{Assumption}

\newtheorem{innercustomthm}{Assumption}
\newenvironment{customthm}[1]
  {\renewcommand\theinnercustomthm{#1}\innercustomthm}
  {\endinnercustomthm}

\newtheorem{theorem}{Theorem}

\newtheorem*{I*}{Invertibility Condition (IC)}
\newtheorem*{CC*}{Compatibility Condition (CC)}
\newtheorem*{CC2*}{Eigenvalue Condition (EC)}
\newtheorem*{CC3*}{Eigen-Compatibility Condition (ECC)}

\newtheorem*{GIR*}{Group Irrepresentable Condition (GIC)}
\newtheorem*{IR*}{Irrepresentable Condition$^*$ (IC$^*$)}
\newtheorem*{BM*}{Beta Min Condition (BM)}
\newtheorem*{SC*}{Substitution Condition (SC)}

\newcounter{defcounter}
\usepackage{xr}
\makeatletter
\newcommand*{\addFileDependency}[1]{
  \typeout{(#1)}
  \@addtofilelist{#1}
  \IfFileExists{#1}{}{\typeout{No file #1.}}
}
\makeatother

\newcommand*{\myexternaldocument}[1]{
    \externaldocument{#1}
    \addFileDependency{#1.tex}
    \addFileDependency{#1.aux}
}

\myexternaldocument{Appendix}

\begin{document}

\title{Heterogeneous Endogenous Effects in Networks \footnote{I would like to thank Francesca Molinari, Matthew Backus, David Easley, Marten Wegkamp, Donald Kenkel, Zhuan Pei, Douglas Miller, Joris Pinkse, Peter Hull, Yanlei Ma and participants in seminars and conferences at which this paper was presented. All remaining errors are mine.}}

\date{\today}

\author{Sida Peng\thanks{Microsoft Research, sidpeng@microsoft.com}
}

\pagenumbering{gobble} 

\maketitle

\hskip 80pt

\begin{abstract}
\footnotesize{

\noindent This paper proposes a new method to identify leaders and followers in a network. Prior works use spatial autoregression models (SARs) which implicitly assume that each individual in the network has the same peer effects on others. Mechanically, they conclude the key player in the network to be the one with the highest centrality. However, when some individuals are more influential than others, centrality may fail to be a good measure. I develop a model that allows for individual-specific endogenous effects and propose a two-stage LASSO procedure to identify influential individuals in a network. Under an assumption of sparsity: only a subset of individuals (which can increase with sample size $n$) is influential, I show that my 2SLSS estimator for individual-specific endogenous effects is consistent and achieves asymptotic normality. I also develop robust inference including uniformly valid confidence intervals. These results also carry through to scenarios where the influential individuals are not sparse. I extend the analysis to allow for multiple types of connections (multiple networks), and I show how to use the sparse group LASSO to detect which of the multiple connection types is more influential. Simulation evidence shows that my estimator has good finite sample performance. I further apply my method to the data in \cite{TheDiffusionofMicrofinance2013} and my proposed procedure is able to identify leaders and effective networks.\\

{\bf key words:} key players, network, endogenous effects, spillovers, high-dimensional models, LASSO, model selection, robust inference
\vfill
}

%
%
\end{abstract}

\clearpage

\pagenumbering{arabic}

\section{Introduction}
\label{sec_introduction}
How an individual's behavior is affected by the behavior of her neighbors in an exogenously given network is an important research question in applied economics. With the increasing availability of detailed data documenting connections among individuals, spatial autoregression models (SARs) have been widely applied in the empirical networks literature to estimate endogenous effects. 

In SARs, an individual's behavior depends on the weighted average of other individuals' behaviors \citep[see][]{Anselin1988, PRUCHA1998}. Standard SARs assume that the peer effects/endogenous effects are the same across individuals in a network. Each individual influences her neighbors \textit{at the same rate} regardless of who she is. However, in many contexts, some individuals are clearly more influential than others. For example, \cite{MasandMoretti} finds that the magnitude of spillovers varies dramatically among workers with different skill levels. \cite{Clark2006} also notes that popular teenagers in a school have much stronger influence on their classmates' smoking decisions than their less popular peers.


I propose a novel SAR model which allows for \emph{heterogeneous} endogenous effects. Each individual in a network simultaneously generates an outcome that takes into account all her neighbors' behaviors. Unlike standard SARs, each individual has an individual-specific effect on her neighbors. As a result, there are as many coefficients for individual-specific endogenous effects as there are individuals in the network. To achieve identification, I assume that ``truly-influential" individuals only constitute a small fraction of the total population. In other words, individual-specific coefficients are assumed to be sparse. This assumption allows me to estimate the model via the least absolute shrinkage and selection operator (LASSO). The LASSO procedure penalizes the $l_1$ norm for the coefficients of heterogeneous endogenous effects. The geometry of the $l_1$ norm enforces the sparsity in the LASSO estimators. If a coefficient is selected by LASSO (i.e. the estimated coefficient is non-zero), the individual associated with this coefficient can influence all her neighbors at her specific rate. Otherwise the LASSO estimator will indicate that the individual has no influence on her neighbors. With some restrictions on the network structures, I show that the LASSO estimates for heterogeneous endogenous effects have near oracle performance \citep[see][]{Peter2011}. In other words, the selection of influential individuals is consistent and the convergence rate of non-zero LASSO estimates is the same as the convergence rate that would have been achieved if the truly influential individuals were known.   


One challenge in my estimation process is the presence of endogeneity in the spatial lag and error term. As with standard SARs, the dependent variable in my model is used to construct spatial lags as an independent variable. As a result, the regressors are correlated with the error term and estimates would be biased if we were to apply LASSO directly.

First I propose a set of novel instruments to address the endogeneity. Following \cite{PRUCHA1998}, I express the dependent variable as an infinite sum of functions consisting of exogenous characteristics and an adjacency matrix. I show that the exogenous characteristics of influential individuals can be used as instruments for their neighbors. Then I design a two-stage estimation process for heterogeneous endogenous effects using LASSO at each stage. \emph{In the first stage}, I use LASSO to estimate the coefficients for the instruments. These estimated coefficients and instruments are then used to create a synthetic dependent variable. \emph{In the second stage}, I replace the dependent variable in the spatial lags with the synthetic variable to perform the LASSO estimation.    

The next challenge is to construct robust confidence intervals for my LASSO type two-stage estimator. As pointed out in \cite{LeebandPotscher2008}, it is impossible to estimate the distribution for post-model-selection estimators. Consistent model selection by LASSO is only guaranteed when all non-zero coefficients are large enough to be distinguished from zero in a finite sample (i.e. a condition usually named ``beta-min"). LASSO may fail to select regressors with very small coefficients, resulting in omitted variable bias in the post LASSO inference. 

I propose a bias correction for my two-stage estimator following the recent LASSO inference literature \citep[see][]{DML2016, Belloni2015, vandeGeer2014, Javanmard, zhangandzhang2011, Zhu2016}. The idea is to correct the first order bias and make the estimators independent from the model selection. Heuristically, shrinkage bias due to the $l_1$ penalty in LASSO can be expressed as a function of the LASSO estimators. Normality can still be achieved after adjusting for this bias. I show that this strategy also works in my two-stage estimation process under the presence of spatial errors. I derive the asymptotic normality for my ``de-bias" two-stage LASSO estimator and conduct robust inference including confidence intervals.  

My model can be extended to allow for more flexible structures of influencing. One real world scenario is a network which consists of many local leaders. Local leaders may only influence individuals within their own cliques/groups but have no influence on individuals outside their cliques/groups. Influence from local leaders can be represented by a homogenous endogenous effects. Exogenous effect and homophily within cliques may also be present.  


To solve the problem in this scenario, I modify my model by bringing back the classical SAR model. I assume that there are both local leaders and global leaders in the network. In contrast to local leaders, global leaders can influence individuals across different cliques. I show that homogeneous endogenous effects across local leaders, exogenous effects, and correlated effects can be identified under assumptions similar to those \cite{Bramoull2009}. Under a sparsity assumption, the endogenous effects of global leaders, whose influence remain individual-specific, can also be identified under the same set of instruments proposed in my main model.  
 
Another real world scenario is the existence of multiple types of connections among individuals. For example, connections among individuals can be classified as social (e.g.  friendship, kinship) or economic (e.g.  lending, employment). In epidemiology, infectious disease can be spread through air, insects, or direct contact. It is important to identify which networks are more efficient at transmitting the endogenous effects.

I model different types of connections as multiple networks. I propose the use of sparse group LASSO to estimate a heterogeneous endogenous effects model with multiple networks. The sparse group LASSO penalizes both the $l_1$ norm and the $l_2$ norm for each coefficient in each type of connection and thus selects both the influential individuals and effective types of connections that generate spillover. I derive the convergence rate and prove the consistency of selection for this estimator. To the best of my knowledge, my paper is the first to show statistical properties for sparse group LASSO.


I provide simulation evidence for networks of different sizes and different data generating algorithms. The empirical coverage of my proposed estimators is close to the nominal level in all scenarios. Similar results are also found in models with multiple networks and with cliques.  

I apply my method to study villagers' decisions to participate in micro-finance programs in rural areas of India as in \cite{TheDiffusionofMicrofinance2013}. Instead of simulating the contagion process and reconstructing the data into panel format as in \cite{TheDiffusionofMicrofinance2013}, my method allows researchers to directly analyze cross-sectional data under the standard equilibrium assumptions. Among different social and economic networks, my method shows that some networks such as ``visit go-come" and ``borrow money", are much more effective at influencing villagers' decisions than other networks such as ``go to temple together" and ``medical help". I further show that individuals in certain careers such as agricultural workers, Anganwadi teachers and small business owners are more likely to influence villagers.

\subsection{Literature Review}
This paper brings together literature on spatial autoregression model, LASSO and networks.

\textbf{SARs}:\\
SARs have been widely applied in empirical studies. For instance, they have been used to study peer effects in labor productivity \citep[see][]{MasandMoretti, Guryan2009, BANDIERA2009}, smoking behavior among teenagers \citep[see][]{Krauth2005, Clark2006, Nakajima2007}, educational achievements among different student groups \citep[see][]{Sacerdote2001, Neidell2010}, systemic risk in finance \citep[see][]{Bonaldi2015, Denbee2015}, and the adoption of new agricultural technologies \citep[see][]{Coelli2002, Conley2010}. My paper proposes a novel extension of standard SARs that could be used to identify influential individuals in a given network.  My methodology for estimating such a model could easily be adopted in existing empirical SARs analyses to identify influential individuals who influence their peers productivity, smoking decisions, or financial holdings.

More specifically, my model extends existing SARs literature by introducing \emph{heterogeneous} endogenous effects. Until very recently, SARs always assume a constant rate of dependence for endogenous effects across different individuals. Moreover, row-sum normalization is widely adopted in the estimation, which may further result misspecification under the existence of heterogeneous endogenous effects. (see  \cite{Spatialautocorrelation}, the first monograph on the topic, and the later studies, \cite{Upton, Anselin1988, Cressie, Lee2010, Lee2010b, Lee2016}). Recent developments in social interaction literature incorporate individual characteristics into SARs, essentially allows for a limited degrees of pre-stipulated heterogeneity, for example, \citep[see][]{ Pinkse2002}. In contrast, this paper shows that heterogeneous endogenous effects can be identified from individuals' outcomes instead of being pre-specified through individuals' characteristics. Another article studying individual heterogeneity from random coefficient aspect is \cite{Masten2018}. \cite{Masten2018}'s model assumes each individual has different rate of receiving the influence while my model assumes each individual has different rate of influencing her neighbors. The two different models yield two different identification strategies. This paper contributes to the literature by showing heterogeneous endogenous effects can be directly identified using cross-sectional data.    
 
To estimate the heterogeneous endogenous effects in my model, I propose a methodology that is different from standard SARs literature. In classic SARs, there is only one endogenous variable and hence it is sufficient to identify the model through only one instrument. In my model, the number of potentially endogenous variables increases as the number of observations increases. As a result, I propose a set of instruments that contain the same number of instruments as the total number of individuals. Each instrument is essentially a decomposition from the standard SARs instruments as in \cite{PRUCHA1998}, \cite{Lee2002}, \cite{Lee2003} and \cite{Lee2004}.   

I show that my model can be combined with SAR model under homogeneous endogenous effects, exogenous effects and correlated effects. To solve the classical ``reflection problem" as in \cite{Manski1993}, I adopt the same strategy as in \cite{Bramoull2009} and show that ``neighbors' neighbor" instruments can be combined with my set of instruments to identify additional structures in the spillovers.    

This paper also contributes to literature that models multiple networks through SARs. In standard SARs, multiple networks are modeled as higher order spatial lags \citep[see][]{Lee2010}. My model allows each individual to have her own specific endogenous effects in each network. This design enables the selection at the network level which implies some networks can be classified as completely irrelevant to decision-making. Furthermore, as the asymptotic allows the number of networks $q$ to go to infinity, my model can handle those cases when researchers observe a large number of networks $(q>n)$.  

\textbf{LASSO}:\\
My paper contributes to the growing literature on endogenous regressors in LASSO estimators.  For instance, \cite{BelloniChernozhukov2013} proposes the double selection mechanism to study confounded treatment effects. \cite{FAN2014} proposes a GMM type estimator to deal with many endogenous regressors. \cite{GautierandTsybakov} proposes a Self Tuning Instrumental Variables (STIV) estimator. The paper that is closest to mine is \cite{Zhu2016}, which studies the statistical properties of two-stage least square procedure with high-dimensional endogenous regressors. The two-stage estimator proposed in \cite{Zhu2016} assumes an i.i.d error term for the first stage. However, this assumption is incompatible with SAR model as the structure assumptions in SAR lead to a first stage with correlated unobservables. In this paper, I derive the rate of convergence and consistency of selection for a two-stage LASSO estimator under the presence of spatial errors. I show that a modified ``de-bias" LASSO estimator accounting for spatial errors can be constructed for my estimator in a manner similar to \cite{zhangandzhang2011}, \cite{Buhlmann2013}, \cite{vandeGeer2014}, and \cite{Zhu2016}. I derive its asymptotic distributions and show how to perform inference.

This paper also extends LASSO literature by deriving statistical bounds and consistency of selection for \textit{sparse group} LASSO estimator. \cite{YuanandLin} proposes the group LASSO, in which explanatory variables are represented by different groups. The group LASSO assumes that sparsity exists only among groups, i.e. some groups of variables are relevant while other groups are not. \cite{Simon2013} proposes the sparse group LASSO, which further allows sparsity within each group, i.e. some regressors within the relevant groups can also be irrelevant.  \cite{Bunea2013} derives statistical properties for the square-root group LASSO, which combines group LASSO and square-root LASSO.  When estimating a heterogeneous endogenous effects model with multiple networks, I establish both statistical bounds and consistency of selection for the \textit{sparse group} LASSO estimator. \textit{Sparse group} LASSO differs from group LASSO by allowing the number of regressors in each group to also goes to infinity as the number of groups goes to infinity. To the best of my knowledge, this paper is the first to show asymptotic statistical properties for the \textit{sparse group} LASSO estimator.

\textbf{Networks}:\\
My paper shares similar microfoundations with SARs as discussed in \cite{Blume2015}, where the individual utility function can be written as a linear summation of the private and social components. The private component is a quadratic loss function on individual's efforts. The social components depend on the network structure as well as the efforts of one's neighbors. While the marginal rate of substitution between the private and social components of utility is assumed fixed in SARs, I assume this rate is individual-specific and depends on one's neighbors. My paper applies and extends LASSO approaches to deal with a high-dimensional problem in networks. The total number of possible edges in a network is $n^2$, however, the social interaction networks we often observe are far more sparse. This is an ideal setting where penalized estimators like LASSO could be applied. \cite{Manresa2013} studies the heterogeneous exogenous effects in a network using LASSO. \cite{dePaula2015} explores the use of LASSO to recover network structures. Both these two papers consider panel data and rely on repeated observations of the same network to identify their models. My model considers cross-sectional data. To identify an individual's endogenous effects, I rely on the variations in her neighbors' outcome. 

My paper also relates to the literature on identifying the key players in the network following \cite{KeyPlayer}, \cite{Antoni2009}, and \cite{Horracea2016}. Under the framework of SARs, every individual is assumed to have the same endogenous effects. As a result, individuals who are well-connected in the network (with high centrality measure) are considered as the key players. However, well connected individuals may effectively have zero effects on their neighbors under heterogeneous endogenous effects. Indeed, as shown in the empirical application, well connected villagers such as tailors, hotel workers, veterans, and barbers are not influential in other villagers' decisions to join the micro-finance program.

The rest of this paper is organized as follows: in Section 2, I introduce the model; in Section 3, I discuss assumptions; in Section 4, I design estimation procedures and derive consistency and asymptotic properties; in Section 5, I show finite sample performance using Monte Carlo simulations; in Section 6, I apply my proposed model to study influential individuals and effective networks in promoting micro-finance programs in rural India; and in Section 7, I conclude.

\section{Models}
\label{sec_model_2}

\subsection{Benchmark Endogenous Effects Model}
Let $n$ denotes the total number of observed individuals in a network. The outcome of individual $i$ is denoted as $d_i$ and is the variable of interest.  Here $d_i$ can represent outcome variable of interest associated with individual $i$, such as whether to join a program or $i$'s labor productivity. In standard SAR model, it is assumed that the outcome of each neighbor of individual $i$ impacts her outcome homogeneously through a constant rate $\lambda_0$: 
\begin{equation}
d_i = \lambda_0 \sum_{j \in N_i} d_j + x_i\beta_0+\epsilon_i,
\end{equation}
where the set $N_i$ is defined as individual $i$'s neighbors. The matrix form of this model is expressed as follows:
\begin{equation}\label{SAR}
D_n = \lambda_0 M_{n}D_n +X_n\beta_0 + \epsilon_n,
\end{equation}
where $D_n = (d_1, d_2, \cdots, d_n)'$ is the $n$-dimensional vector of observable outcomes. The $n$ by $k$ matrix $X_n$ represents the observable exogenous characteristics of individuals. When $\epsilon_n$ is specified as an $n$-dimensional vector of independent and identically distributed disturbances with zero mean and a constant variance $\sigma^2$, equation (\ref{SAR}) is also called a mixed regression model. 

The spatial weight matrix $M_n$ is of size $n$ by $n$, where the $(i,j)$th entry represents the connection between individual $i$ and individual $j$. In empirical studies, the spatial weight matrix is often replaced by the adjacency matrix \citep[see][]{Andreas2009, AcemogluGarciaJimenoandRobison2012, TheDiffusionofMicrofinance2013}: the $(i,j)$-th entry of the matrix $M_n$ takes value $1$ if individual $i$ and individual $j$ are connected and takes value $0$ otherwise; the diagonal entries of the matrix $M_n$ are always $0$s. In the SAR literature, spatial weight matrix or adjacency matrix is taken as exogenous. My method follows this assumption and is designed for cross-sectional data. However, it is important to recognize that social networks do change over time and it is then important to take network formation into modeling (see \cite{networformation}, \cite{sheng2016}).    

In mixed regression model, endogenous effects \citep[see][]{Manski1993} or network effects \citep[see][]{Bramoull2009} are captured by the scalar $\lambda_0$. An implicit assumption in equation (\ref{SAR}) is that $\lambda_0$, the rate of endogenous effects, is identical across all individuals in the network. Although a limited degree of heterogeneity can be built into the adjacency matrix via pre-specified structure assumptions, the identification potential for heterogeneous endogenous effects has not been fully explored.   This limitation has been noted in various studies \citep[see][]{Andreas2009, dePaula2015}.  I relax this assumption by proposing a more flexible model that allows and identifies individual-specific endogenous effects as discussed below. 
%
%
%
%

\subsection{Heterogeneous Endogenous Effects Model}
I propose the following model: 
\begin{equation}\label{HEEMreduce}
d_i = \sum_{j \in N_i} d_j\eta_j + x_i\beta+\epsilon_i
\end{equation}
where $N_i$ represents the set of individual $i$'s neighbors and $\eta_j$ represents the endogenous effects of individual $j$ on the outcome of all her neighbors $i \in N_j$. the model can be rewritten in matrix form as:
\begin{equation}\label{HEEM}
D_n = \Big(M_n \circ D_n\Big)\eta_0 + X_n\beta_0 + \epsilon_n,
\end{equation}
where $\eta_0 = (\eta_1, \eta_2, \cdots, \eta_n)'$ is a vector of parameter of size $n$ by 1. The $i$th entry in $\eta_0$ represents the endogenous effects of individual $i$ on her neighbors. This model allows for individual heterogeneity to interact with endogenous effects so that every individual is allowed to have her own coefficient $\eta_i$. My model allows some $\eta_j = 0$. In other word, there are individuals that impose no endogenous effects on their neighbors. I define those individuals with $\eta_j \neq 0$ as influential.

The operator $\circ$ is defined between a $n$ by $n$ matrix $M_n$ and a $n$ by 1 vector $D_n$ as 
\begin{equation*}
M_n \circ D_n  = M_n \cdot \text{ diag}(D_n) = C,
\end{equation*}
where $diag(\cdot)$ is the diagonalization operator and $C_{i, j} = M_{i, j}d_{j}$.

Note that in contrast to fixed rate $\lambda_0$ specified in equation (\ref{SAR}), even though each neighbor of individual $j$ is assumed to receive the same influence $d_j\eta_j$ from her, each individual is allowed to influence her neighbors at her own rate $\eta_j$. 

A more generalized form of the model is to replace $\eta_{j}$ with $\eta_{ij}$. This generalization can further capture the different perceiving rate of endogenous effects on the same leader but among different followers. However, the number of parameters increase from $n$ to $n^2$ and the model becomes too saturated to estimate.     

Another direction of modeling is to assume heterogeneous perceiving rate of endogenous effects but maintaining the assumption of homogeneous influencing rate, for example

\begin{equation*}
d_i = \lambda_{i}^*\sum_{j \in N_i} d_j + x_i\beta+\epsilon_i
\end{equation*}

This model has been studied using random coefficients approach in \cite{Masten2018}. This specification differs from equation \eqref{HEEMreduce} as it prevents any sparsity pattern on $\lambda_{i}^*$ to be passed to a first stage regression as shown in proposition \ref{proposition1}. As a result, the identification strategies are fundamentally different.

Equation (\ref{HEEM}) can be derived from a bayesian Nash Equilibrium. Let $(x_i, \epsilon_i)$ denotes an individual's type, where $x_i$ is publicly observed characteristics and $\epsilon_i$ is private characteristics only observable by $i$. Individual $i$'s utility depends on her own action and characteristics as well as her neighbors' actions. Individual $i$ chooses action $d_i$ to maximize the following utility: 
\begin{equation*}
U_i(d_i,d_{-i}) = (x_i\beta+\epsilon_i)d_i-\frac{1}{2}d_i^2  + \sum_{j \in N_i}d_jd_i\eta_j
\end{equation*}
The first order condition yields equation (\ref{HEEM}). The micro-foundations derived above is similar to the one for SARs as discussed in \cite{Blume2015}.


\subsection{Motivating Examples}
\textbf{Peer Effects in Labor Productivity:}

Understanding the mechanism and magnitude of the dependence of labor productivity on coworkers is an important question for economists and policy makers. As found in \cite{MasandMoretti}, workers respond more to the presence of coworkers with whom they frequently interact. Another modern example is on code sharing platforms like GitHub. Programmers provide efforts to an open source project depending on how much other programers are contributing. In such cases, the influence level of each individual to hers coworkers is not the same. Equation (\ref{HEEM}) can be used to incorporate such differences.
\begin{equation*}
y_i = \sum_{j \in N_i} y_j\eta_j + x_i\beta+\epsilon_i,
\end{equation*}
where $y_i$ is individual $i$'s productivity, and $\eta_j$ represents the size of influence of coworker $j$ -- all else being equal, the additional effect on individual $i$'s productivity if individual $j$ becomes her coworker. 

Note that if we restrict the parameters $\eta_j$ to be the same across different workers, then we are back to the classical SARs setting as laid out in equation (\ref{SAR}). Thus, $\lambda = \frac{1}{n}\sum_{j = 1}^n\eta_j$ can be interpreted as the averaged spillover effects in the canonical sense. Notice that $\lambda$ converges to 0 when the spillover effects are sparse and it may lead to the conclusion that there is no spillover in the network under such scenario.    

Define 
\begin{equation*}
\lambda^*  = \frac{1}{\sum \mathbf{1}_{\eta_j \neq 0}}\sum_{j = 1}^n\eta_j \mathbf{1}_{\eta_j \neq 0} 
\end{equation*}
as the averaged endogenous effects for influential workers. The estimand $\lambda^*$ is arguably better than $\lambda$ for average endogenous effects as it excludes individuals who are not influential to their neighbors. 

\textbf{Online Opinion Leaders:}

A decision can represent whether to ``tweet" a news story seen online. When individuals make such decisions, they are often influenced by several online opinion leaders -- whether those people also ``tweet" the news or not. Political figures may have stronger influence on people's tweet for political news than celebrities. And vice versa for entertainment news. Assume a binary decision $(0,1)$ is made from a bayesian Nash Equilibrium, such that   
\begin{equation*}
d_i^* = \sum_{j \in N_i} d_j^*\eta_j + x_i\beta+\epsilon_i,
\end{equation*}
where $d_i^*$ is the probability of individual $i$ playing action 1, and $\sum_{j \in N_j} d_j^*\eta_j$ is the expected endogenous effects from $i$'s neighbors $N_i$ . Define

\begin{equation*}
S = \{j: \eta_j \neq 0\}
\end{equation*}
as truly influential opinion leaders. My method provides a way to estimate $\hat S$ which is asymptotically consistent with $S$. This is an important metric to policy makers or private sectors as targeting or nudging the influential individuals is usually more efficient than targeting the entire population. 

\subsection{Two Extensions of the Model}
\textbf{Heterogeneous Endogenous Effects Model with Cliques:}

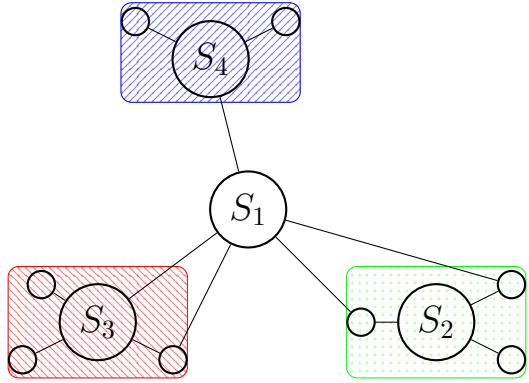
\begin{figure}[H]
\begin{center}
\begin{tikzpicture}
  [scale=.5,auto=left,every node/.style={circle}]
  \node[main node] (n1) at (9,3) {$S_1$};
  \node[main node] (n4) at (14,0) {$S_2$};
  \node[main node] (n9) at (5,0) {$S_3$};
  \node[main node] (n12) at (8,7) {$S_4$};
  \node[main node] (n2) at (3.5,1)  {};
  \node[main node] (n5) at (3,-1)  {};
  \node[main node] (n10) at (7,-1)  {};
  \node[main node] (n3) at (12,0)  {};
  \node[main node] (n6) at (16,1)  {};
  \node[main node] (n11) at (16,-1)  {};
  \node[main node] (n13) at (6,8)  {};
  \node[main node] (n14) at (10,8)  {};

  \begin{scope}[on background layer]
\node[fit=(n2) (n5) (n10), rectangle, rounded corners, pattern=north west lines, pattern color=red, draw=red, fill opacity=0.5,, scale=0.9] () {};
\node[fit=(n3) (n6) (n11), rectangle, rounded corners, pattern=dots, pattern color=green, draw=green, fill opacity=0.5, , scale=0.9] () {};
\node[fit=(n12) (n13) (n14), rectangle, rounded corners, pattern=north east lines, pattern color=blue, draw=blue, fill opacity=0.5, , scale=0.9] () {};
\end{scope}
  \foreach \from/\to in {n1/n9,n1/n3,n4/n3, n1/n6, n1/n10,n9/n10,n4/n11, n4/n6, n9/n2,n9/n5,n12/n13,n12/n14,n1/n12}
    \draw (\from) -- (\to);
\end{tikzpicture}
\end{center}
\caption{Local Leader}
\label{fig1}
\end{figure}

There are two important assumptions for equation \eqref{HEEM}. First is the sparsity assumption, which requires non-influential individuals to have completely zero influence on their neighbors. Second it assumes away the exogenous and correlated effects. Consider a network composed of many cliques (small groups of connected individuals). Each clique has its local leaders who only influence individuals within their own clique. Figure \ref{fig1} provides an example of such a network structure.  Note that in Figure \ref{fig1}, node $S_2$, $S_3$ and $S_4$ represent local leaders who only influence individuals within their own cliques. On the contrary, node $S_1$ represents a global leader who can influence individuals across different cliques. For example, one can think about the local leaders $S_2$, $S_3$ and $S_4$ as local news channels while $S_1$ is the national news channel. Furthermore, within each clique, there might exist exogenous and correlated effects. I propose an extension to my heterogeneous endogenous effects model which could address such concerns. I assume that all local leaders will influence their followers at a small but similar rate while global leaders can have different effects on their audience.   

First, I consider the following extension to introduce only homogeneous endogenous effects:
\begin{equation*}
d_i = \sum_{j \in N_i} d_j\eta_j + \gamma_0 \sum_{j \in N_i} d_j + x_i\beta_0+\epsilon_i,
\end{equation*}
which can be represented in matrix form as:
\begin{equation}\label{HEEMwHE}
D_n = \Big(M_n \circ D_n\Big)\eta_0 +  M_n D_n\gamma_0 + X_n\beta_0 + \epsilon_n,
\end{equation}
where $\eta_0' = (\eta_1, \eta_2, \cdots, \eta_n)'$. The new term $\gamma_0\sum_{j \in N_i} d_j$ captures influence from the local level. Note that this is the same term as the spatial lag in the benchmark spatial autoregression model. The vector $\eta_0$ captures the heterogeneous endogenous effects of global leaders. 


To further include exogenous and correlated effects, consider the following model:
\begin{equation}\label{HEEMwHEEC}
d_i = \sum_{j \in N_i} d_j\eta_j + \gamma_0^{end} \sum_{j \in N_i} d_j + \gamma_0^{exo} \sum_{j \in N_i} x_j + x_i\beta_0+\mu_c+\epsilon_i,
\end{equation}
Besides the first term representing heterogeneous endogenous effects, equation \eqref{HEEMwHEEC} is the same as the model in \cite{Manski1993}. 

\textbf{Heterogeneous Endogenous Effects Model with Multiple Networks:}

Individuals are often connected with each other through more than one type of network. For example, one's financial network (for borrowing/lending) may be different from hers relative network or even friendship network. A common strategy in empirical application is to ``pool" the networks by mixing all types of connections into one network. This approach may increase the noise in the network measurement when the outcome variables only depend on certain types of networks.   

To capture different types of connections among the same set of individuals, we can incorporate multiple networks in the heterogeneous endogenous model. More specifically, a separate adjacency matrix can be constructed for each type of network. For instance, the $(i,j)$-th entry of the adjacency matrix representing friendship takes value $1$ if individual $i$ and individual $j$ are friends and takes value $0$ otherwise; that representing the borrowing/lending network takes value $1$ if individual $i$ and individual $j$ lend money to each other and takes value $0$ otherwise. 

Let $q$ be the total number of different types of networks. Define $M_n^l$ as the adjacency matrix for the $l$th network. The heterogeneous endogenous effects model with multiple networks is defined as
\begin{equation}\label{HEEMwMN1}
d_i = \sum_{l=1}^q\sum_{k \in N_i} d_k^l\eta_k^l + x_i\beta_0+\epsilon_i
\end{equation}
Note that in this model, different networks could potentially bear different endogenous effects for the same individual. In equation (\ref{HEEMwMN1}), coefficient $\eta_k^l$ represents the rate of endogenous effect of individual $k$ through network $l$. As a result, we have $nq+k$ coefficients for endogenous effects. In addition, I assume endogenous effects from different types of networks are linearly additive. The model can also be rewritten in matrix form as:  
\begin{equation}\label{HEEMwMN}
D_n = \sum_{l = 1}^q \left(M_n^l \circ D_n \right) \eta^l_0 + X_n\beta_0 +\epsilon_n,
\end{equation}
where $M_n^l$ is the adjacency matrix for network $l$. $\eta^l = (\eta_1^l, \eta_2^l, \cdots, \eta_n^l)'$ is an $n$ by 1 vector for $l = 1, 2, \cdots, q$. Define a network $l$ as efficient network if $\eta_i^l \neq 0$ for at least one individual $i = 1, 2, \cdots, n$. In this specification, leaders can only influence their neighbors through efficient networks and non-efficient networks are completely independent from the outcome variable.

\section{Assumptions}
\label{sec_identification}
The assumptions discussed in this section combine both standard SARs type assumptions and LASSO type assumptions. SARs type assumptions ensure the existence of valid instruments. LASSO type assumptions provide sufficient and necessary conditions for valid inference. The identification for influential individuals is achieved through its coincidence with the sparsity pattern in the reduced form estimator.

First recall the benchmark SAR model:
\begin{equation}\label{SAR2}
D_n = \lambda_0 M_{n}D_n +X_n\beta_0 + \epsilon_n,
\end{equation}
By rearranging the above equation, we can express endogenous variable $M_nD_n$ solely as a function of $X_n$ and $M_n$, since:
\begin{equation*}
D_n = J_n^{-1}X_n \beta_0 + J_n^{-1} \epsilon_n
\end{equation*}
where $I_n$ is the $n$ by $n$ identity matrix and $J_n = I_n - \lambda_0 M_n$.  It is straightforward that $J_n^{-1}X_n$ can serve as valid instruments for $M_nD_n$. As a result, the identification and estimation of equation (\ref{SAR2}) can be achieved through either 2SLS or GMM as proposed in papers such as \cite{Prucha1995}, \cite{PRUCHA1998} \cite{Lee2002}, \cite{Lee2003}, and \cite{Lee2004}. Following the same strategy, I derive a set of instruments by solving $D_n$ as a function of exogenous variables and the adjacency matrix.

\subsection{Assumptions for the Heterogeneous Endogenous Effects Model}
Without additional restrictions, equation \eqref{HEEM} could not be estimated through canonical method as the number of parameters $n+k$ is greater than the number of observations $n$.  

\begin{assumption}[Sparsity]\label{assumption1}
Let $S_n \subset \{1, 2, \cdots, n\}$ denotes the set of influential individuals (i.e. $\eta_j \neq 0$). Let $s_n = |S_n|$ be the number of elements in $S_n$. 
\begin{equation*}
s_n = o\left(\frac{\sqrt{n}}{\log n}\right), \quad \text{  as  } n \rightarrow \infty
\end{equation*}
\end{assumption}
Assumption 1 is usually referred to as ``sparsity" assumption. The assumption that most individuals in a network are not influential is plausible under many circumstances. For example, opinion leaders on social media only constitute a very small fraction of internet users. Star programmers who can encourage other programmers to work with them on Github are also a small portion of all Github users. On the other hand, assumption \ref{assumption1} can be easily violated when there exists many local leaders. For example, when studying the peer effects in obesity among school children, each class/grade can be treated as a clique and as the number of cliques increases, the number of leaders may increase at a rate of $O(n)$. In next subsection, I propose an extension of the current model to partially address this problem.   

\begin{assumption}[SAR restrictions]\label{assumption2}
\leavevmode
\begin{itemize}
\item[-] There exists an $\eta_{\max}<1$ such that $\|\eta_0\|_\infty \leq \eta_{max}$
\item[-] The $\epsilon_j$ are i.i.d sub-Gaussian random variable with 0 mean and variance $\sigma^2$
\item[-] The regressors $x_i$ in $X_n$ are non-stochastic and uniformly bounded for all n. $\lim_{n \rightarrow \infty}X_n'X_n/n$ exists and is nonsingular
\item[-] The minimum eigenvalue of $(I_n-M_n\circ\eta_0)$, $\Lambda_{min}$, is uniformly bounded away from 0.  
\end{itemize}
\end{assumption}
The restriction on $\eta_0$ excludes the unit root process and ensures the uniqueness of equilibrium. It guarantees the invertibility of $\Big(I_n-M_n \circ \eta_0\Big)$. This is a fundamental assumption required in all SAR literatures see \cite{Upton, Anselin1988, Lee2016}). 

The assumption on the error term currently excludes exogenous effects and correlated effects from my model. The SAR model with this exclusion restriction is known as mixed regression model as in \citep[see][]{Lee2002}. The main challenge to relax this assumption is known as ``reflection problem" as pointed out in \cite{Manski1993}. I adopt a strategy similar to \cite{Bramoull2009} to incorporate both exogenous and correlated effects in next subsection. I also require the error term to be sub-Gaussian so that concentration inequalities can be derived to bound the empirical process. Sub-Gaussian process is known to have ``almost" bounded support due to the fast decay of its tails. This assumption is usually required for high dimensional estimators as in \cite{BelloniChernozhukov2013}, \cite{Zhu2016} and etc.

The assumption on the regressors follows the convention in the SAR literatures see \cite{Anselin1988, Lee2016}. This deterministic design assumption can be extended to the random design cases under my model. In the following context, I focus on the case where $X_n$ is an $n$ by 1 vector and study identification as in \cite{Bramoull2009}. 

I also require the minimum eigenvalue of $(I_n-M_n\circ\eta_0)$ to be uniformly bounded away from 0. This is to prevent the spatial errors from accumulating too fast. The spatial error term after the matrix inversion is $(I_n-M_n\circ\eta_0)^{-1}\epsilon$. A similar assumption can be found in \cite{Prucha1995} and \cite{PRUCHA1998}, which requires the uniformly boundedness of $(I_n-M_n\circ\eta_0)^{-1}$.     


To proceed, recall the definition of the operator ``$\circ$" as $M_n \circ D_n = M_n \cdot \text{ diag}(D_n)$, where diag$(\cdot)$ is the diagonalization operator. Note the following property of the ``$\circ$":
\begin{equation*}
\Big(M_n \circ D_n\Big)\eta_0 = \Big(M_n \circ \eta_0\Big)D_n
\end{equation*}
 If the invertibility of $\Big(I_n-M_n \circ \eta_0\Big)$ is guaranteed, then 
\begin{equation}\label{10}
\begin{split}
& D_n = \Big(M_n \circ D_n\Big)\eta_0 + X_n\beta_0 + \epsilon_n \Leftrightarrow D_n = \sum_{i  =0}^\infty\Big(M_n \circ \eta_0\Big)^i(X_n\beta_0 + \epsilon_n)
\end{split}
\end{equation}

Since $\Big(M_n \circ D_n\Big)\eta_0$ is correlated with $\epsilon_n$ and $\eta_0$ is sparse (i.e. having at most $s_n$ non-zero elements), we need at least $s_n$ instruments to deal with the endogeneity in the model. Using equation (\ref{10}), we can express the expectation of $D_n$ as follows:
\begin{equation}\label{inst}
\begin{split}
E(D_n) =  X_n \beta_0 + \Big(M_n \circ X_n\Big)(\beta\eta_0) + \sum_{i  =2}^\infty\Big(M_n \circ \eta_0\Big)^i X_n\beta_0,
\end{split}
\end{equation}
Let $(\cdot)_S$ denote the operator such that $(M_n)_S$ is a sub matrix of $M_n$ with its columns restricted to columns corresponding to the elements of $S$. The first and second terms of equation (\ref{inst}) suggest that $X_n$ and $(M_n \circ X_n)_S$  can serve as valid instruments to point identify $\beta_0$ and $\eta_0$. 

\begin{proposition}[First Stage Equivalence]\label{proposition1}
Under assumption \ref{assumption1} and \ref{assumption2} 
\begin{equation}\label{12}
\begin{split}
E(D_n)  = X_n \beta_0 + \Big(M_n \circ X_n\Big)\tilde{\eta},
\end{split}
\end{equation}
where $\tilde{\eta}_j = \eta_j \tilde{g}(\eta_0, \beta_0, X_n, M_n)$ for some function $\tilde{g}$ depends on $\eta_0$, $\beta_0$, $X_n$, and $M_n$.
\end{proposition}

Proposition \ref{proposition1} shows the sparsity pattern is preserved when solving the simultaneous equations since $\tilde{\eta}_j = 0$ as long as $\eta_j = 0$ \footnote{\cite{dePaula2015} noticed that sparsity pattern will generally not be preserved during matrix inversion when there is no further restriction on the adjacency matrix. Proposition 1 shows that with pre-existing network structure and homogeneous influence assumption for a given influential individual, sparsity pattern can still be preserved after matrix inversion.}. As a result, the sparsity assumption is also satisfied in equation (\ref{12}), and I can thus estimate equation (\ref{12}) as the first stage using a LASSO type estimator.

Define $W_n$ as the projection matrix on to the  orthogonal space of $X_n$: $W_n = I_n - X_n(X_n'X_n)^{-1}X_n'$  

\begin{figure}[H]
\begin{center}
\begin{tikzpicture}
  [scale=.70,auto=left,every node/.style={circle}]
  \node[main node] (n1) at (5,2) {$S_1$};
  \node[main node] (n4) at (8,-2) {$S_2$};
  \node[main node] (n2) at (4,0)  {};
  \node[main node] (n3) at (9,0)  {};
  \node[main node] (n5) at (3,0)  {};
  \node[main node] (n6) at (10,0)  {};
  \node[main node] (n7) at (5,0)  {};
  \node[main node] (n8) at (8,0)  {};
  \node at ($(n2)!.5!(n3)$) {\ldots};
  \foreach \from/\to in {n1/n2,n4/n2,n1/n3,n4/n3, n1/n5,n1/n6,n4/n5,n4/n6, n1/n7,n1/n8,n4/n7,n4/n8}
    \draw (\from) -- (\to);
        \node (n9) at (6,-3.5)
        {
            $(1)$
        };
\end{tikzpicture}
\hspace{3cm}
\begin{tikzpicture}
  [scale=.70,auto=left,every node/.style={circle}]
  \node[main node] (n1) at (6.5,2) {$S_1$};
  \node[main node] (n4) at (8,-2) {$S_2$};
  \node[main node] (n9) at (5,-2) {$S_3$};
  \node[main node] (n2) at (4,0)  {};
  \node[main node] (n3) at (9,0)  {};
  \node[main node] (n5) at (3,0)  {};
  \node[main node] (n10) at (6,0)  {};
  \node[main node] (n6) at (10,0)  {};
  \node[main node] (n11) at (7,0)  {};
  \node at ($(n2)!.5!(n10)$) {\ldots};
  \node at ($(n3)!.5!(n11)$) {\ldots};
  \begin{scope}[on background layer]
\node[fit=(n2) (n5) (n10), rectangle, rounded corners, pattern=north west lines, pattern color=red, draw=red, fill opacity=0.5, scale=0.9] () {};
\node[fit=(n3) (n6) (n11), rectangle, rounded corners, pattern=dots, pattern color=green, draw=green, fill opacity=0.5,scale=0.9] () {};
\end{scope}
  \foreach \from/\to in {n1/n2,n1/n3,n4/n3, n1/n5,n1/n6, n1/n10,n1/n11,n9/n10,n4/n11, n4/n6, n9/n2,n9/n5}
    \draw (\from) -- (\to);
        \node (n9) at (7,-3.5)
        {
            $(2)$
        };
\end{tikzpicture}
\caption{Examples of networks which violate assumption \ref{assumption3}}
\medskip
\begin{minipage}{0.9\textwidth} 
{\footnotesize
(1) Two influential individual $S_1$ and $S_2$ share the exact same neighbors. (2) The neighbors of an influential individual $S_1$ is a linear combination of $S_2$ and $S_3$'s neighbors. 
}
\end{minipage}
\label{fig2}
\end{center}
\end{figure}
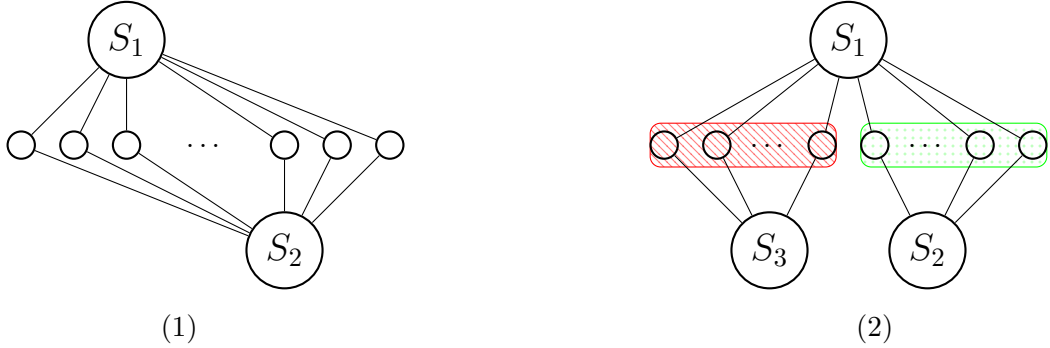

\begin{assumption} [Independence] \label{assumption3}
$W_n(M_n \circ X_n)_S$ has full column rank.
\end{assumption}
The linear independence among $(M_n \circ X_n)_S$ requires the assumption that any two influential individuals may not necessarily connect with identical neighbors. Moreover, assumption \ref{assumption3} also requires that neighbors of an influential individual cannot be a linear combination of neighbors of several other influential individuals, which rules out network structures as depicted in Figure \ref{fig2}. 

\begin{figure}
\begin{center}
\begin{tikzpicture}
  [scale=.70,auto=left,every node/.style={circle}]
  \node[main node] (n1) at (7,2) {$S_1$};
  \node[main node] (n4) at (7,-2) {$S_2$};
  \node[main node] (n2) at (2,0)  {};
  \node[main node] (n3) at (7,0)  {};
  \node[main node] (n5) at (1,0)  {};
  \node[main node] (n10) at (4,0)  {};
  \node[main node] (n6) at (8,0)  {};
  \node[main node] (n11) at (5,0)  {};
  \node[main node] (n12) at (9,0)  {};
  \node[main node] (n13) at (10,0)  {};
  \node[main node] (n14) at (12,0)  {};
  \node[main node] (n15) at (13,0)  {};
  \node[main node] (n16) at (14,0)  {};
  \node[main node] (n17) at (16,0)  {};
  \node at ($(n2)!.5!(n10)$) {\ldots};
  \node at ($(n3)!.5!(n11)$) {\ldots};
  \node at ($(n13)!.5!(n14)$) {\ldots};
  \node at ($(n16)!.5!(n17)$) {\ldots};
  \begin{scope}[on background layer]
\node[fit=(n2) (n5) (n10), rectangle, rounded corners, pattern=north west lines, pattern color=red, draw=red, fill opacity=0.5,, scale=0.9] () {};
\node[fit=(n3) (n6) (n11), rectangle, rounded corners, pattern=dots, pattern color=green, draw=green, fill opacity=0.5, , scale=0.9] () {};
\node[fit=(n12) (n13) (n14), rectangle, rounded corners, pattern=north east lines, pattern color=blue, draw=blue, fill opacity=0.5, , scale=0.9] () {};
\node[fit=(n15) (n16) (n17), rectangle, rounded corners, fill=black!30, draw=black, fill opacity=0.5, , scale=0.9] () {};
\end{scope}
  \foreach \from/\to in {n1/n2,n1/n3,n4/n3, n1/n5,n1/n6, n1/n10,n1/n11,n4/n11, n4/n6, n4/n12,n4/n13, n4/n14}
    \draw (\from) -- (\to);
\end{tikzpicture}
\caption{Examples of networks which satisfies assumption \ref{assumption3}}
\medskip
\begin{minipage}{0.9\textwidth} 
{\footnotesize
The influence of $S_1$ can be identified by comparing red (right shaded) and grey groups (plain), while the influence of $S_2$ can be identified by comparing blue (left shaded) and grey (plain) groups. Or the influence of $S_1$ can be identified by comparing green (dotted) and blue (left shaded) groups, while the influence of $S_2$ can be identified by comparing red (right shaded) and green (dotted) groups.
}
\end{minipage}
\label{fig3}
\end{center}
\end{figure}
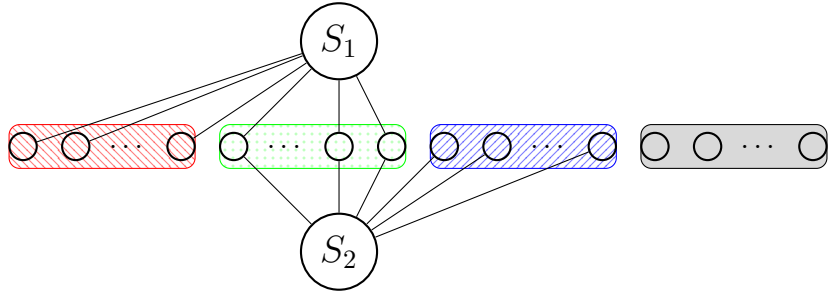

\vspace{-0.5cm}
In other words, as long as each influential individual has a neighbor that is not connected with any other influential individuals, assumption \ref{assumption3} is satisfied. One can think of the identification here as estimating fixed effects from influential individuals as illustrated in Figure \ref{fig3}. Collinearity arises when the fixed effects of two influential individuals are imposed on exactly the same observations.  

Assumption \ref{assumption3} is essentially a restriction on the topology of network structures. It rules out cases like complete network or cases (1) and (2) as in Figure \ref{fig2}. To achieve identification under a cross-sectional network data, one has to rely on certain network structures. Similar assumptions can be found in \cite{Bramoull2009} and \cite{dePaula2015}.      

At this point, if the truly influential individuals set $S_n$ were available to us, we would be able to estimate the model using 2SLS method or GMM. However, in most cases, $S_n$ is not known beforehand. Notice that the identification for the set $S_n$ in the structure model \eqref{HEEM} coincides with the sparsity pattern in the reduced form \eqref{12}. I propose to use a LASSO type estimator to both recover the set of influential individuals and estimate the model. For LASSO to achieve correct recovery, I need the following assumptions:

\begin{assumption}\label{assumption4}[LASSO restrictions]
\hfill\\
{(Irrepresentable Condition)} There is a $\vartheta \in (0,1)$ such that
\[\max_{\|u\|_\infty \leq 1}\left\|diag(f_{S^c}) \cdot \Sigma_{2,1}^M(\Sigma_{1,1}^{M})^{-1}\cdot diag(f_{S})^{-1}\cdot u\right\|_\infty < \vartheta\]

where 
\begin{equation*}
\Sigma := \frac{1}{n}M_n' W_nM_n =  \left( \begin{array}{cc}
\Sigma_{1,1}^M & \Sigma_{1,2}^M \\
\Sigma_{2,1}^M & \Sigma_{2,2}^M \end{array} \right)
\end{equation*}
and
\[f = (I - M_n\circ \eta_0)^{-1}X_n\beta_0\]
{(Beta Min Condition)} There exists $N \in \mathbb{N}$ and a $m >0$ such that $\forall n \geq N$, 
\begin{equation*}
\min(|\eta_{0}|)_S \geq m/\sqrt{n},
\end{equation*}
\end{assumption}

Here define the operator $(\cdot)_S$ as the sub-matrix/vector restricted to the columns/entries corresponding to influential individuals. Similarly, $(\cdot)_{S^c}$ represents the sub-matrix/vector restricted by the columns/entries corresponding to non-influential individuals. Also notice that the invertibility of $\Sigma_{1,1}^M$ is guaranteed by assumption \ref{assumption3}.

\begin{figure}[H]
\begin{center}
\begin{tikzpicture}
  [scale=.70,auto=left,every node/.style={circle}]
  \node[main node] (n1) at (5,2) {$S_1$};
  \node[main node] (n4) at (8,-2) {};
  \node[main node] (n2) at (4,0)  {};
  \node[main node] (n3) at (9,0)  {};
  \node[main node] (n5) at (3,0)  {};
  \node[main node] (n6) at (10,0)  {};
  \node[main node] (n7) at (5,0)  {};
  \node[main node] (n8) at (8,0)  {};
  \node at ($(n2)!.5!(n3)$) {\ldots};
  \foreach \from/\to in {n1/n2,n4/n2,n1/n3,n4/n3, n1/n5,n1/n6,n4/n5,n4/n6, n1/n7,n1/n8,n4/n7,n4/n8}
    \draw (\from) -- (\to);
\node (n9) at (8.8,-2.3)
        {
            $s_4$
        };
\node (n10) at (6,-3.5)
        {
            $(1)$
        };
\end{tikzpicture}
\hspace{3cm}
\begin{tikzpicture}
  [scale=.70,auto=left,every node/.style={circle}]
  \node[main node] (n1) at (6.5,2) {$S_1$};
  \node[main node] (n4) at (8,-2) {};
  \node[main node] (n2) at (4,0)  {};
  \node[main node] (n3) at (9,0)  {};
  \node[main node] (n5) at (3,0)  {};
  \node[main node] (n10) at (6,0)  {};
  \node[main node] (n6) at (10,0)  {};
  \node[main node] (n11) at (7,0)  {};
   \node[main node] (n12) at (3,-2)  {};
  \node[main node] (n13) at (4,-2)  {};
  \node[main node] (n14) at (6,-2)  {};
  \node at ($(n2)!.5!(n10)$) {\ldots};
  \node at ($(n3)!.5!(n11)$) {\ldots};
  \node at ($(n13)!.5!(n14)$) {\ldots};
  \begin{scope}[on background layer]
\node[fit=(n2) (n5) (n10), rectangle, rounded corners, pattern=north east lines, pattern color=blue, draw=blue, fill opacity=0.5, , scale=0.9] () {};
\node[fit=(n3) (n6) (n11), rectangle, rounded corners, pattern=dots, pattern color=green, draw=green, fill opacity=0.5, , scale=0.9] () {};
\node[fit=(n12) (n13) (n14), rectangle, rounded corners, pattern=north west lines, pattern color=red, draw=red, fill opacity=0.5,, scale=0.9] () {};
\end{scope}
  \foreach \from/\to in {n1/n2,n1/n3,n4/n3, n1/n5,n1/n6, n1/n10,n1/n11,n4/n11, n4/n6}
    \draw (\from) -- (\to);
\node (n9) at (8.8,-2.3)
        {
            $s_4$
        };
\node (n10) at (7,-3.5)
        {
            $(2)$
        };
\end{tikzpicture}
\caption{Examples of networks which violate and satisfy assumption \ref{assumption4}}
\medskip
\begin{minipage}{0.9\textwidth} 
{\footnotesize
In (1), the influence of $S_1$ can not be separately identified from $S_4$ although $S_1$ is influential and $S_4$ is non-influential.  In (2), the influence of $S_1$ can be identified by comparing red (right shaded) and blue (left shaded) groups. And then we can find out $S_4$ is non-influential by comparing blue (left shaded) and green (dotted) groups. 
}
\end{minipage}
\label{fig4}
\end{center}
\end{figure}
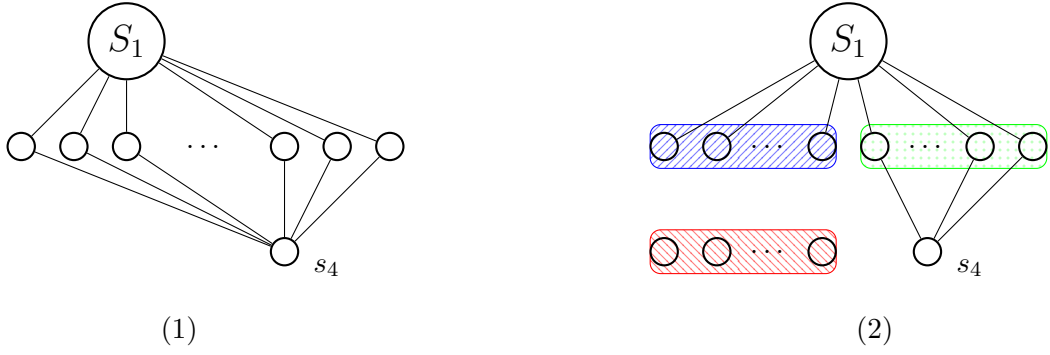

We prove in theorem \ref{theorem2} that assumption \ref{assumption4} is a sufficient condition for the LASSO estimator to achieve a consistent selection for the set $S_n$ in the second stage. While assumption \ref{assumption3} restricts how influential individual may connect with their neighbors,  assumption \ref{assumption4} restricts how non-influential individual may connect. For example, Irrepresentable Condition prevents the neighbors of a non-influential individual to be exactly the same as the neighbors of any influential individual. This is because when two individuals connect with exactly the same neighbors, we cannot distinguish which individual is the true source of influence. This is illustrated in Figure \ref{fig4} (1). On the other hand, assumption \ref{assumption4} does not require full independence between influential individuals and non-influential individuals. This is illustrated in Figure \ref{fig4} (2).

\vspace{-0.5cm}
The Beta Min Condition requires the magnitude of the endogenous effects to be sufficiently strong in order to be detected by LASSO. For example, there does not exist a sequence of individuals whose influence decay to 0 faster than the rate of $1/\sqrt{n}$. Beta Min Condition is restrictive such that it rules out uniform inference and creates additional problems when constructing confidence intervals.      

Equation \eqref{HEEM} can still be consistently estimated under weaker conditions than Irrepresentable Condition and Beta Min Condition. The stronger version is assumed above to ensure selection consistency. As shown in \cite{ZhaoandYu2006}, the Irrepresentable Condition together with the Beta Min Condition are necessary and sufficient conditions for LASSO to achieve consistent model selection.  

The following compatibility condition will guarantee the valid inference using ``de-bias" estimator proposed in the next section.  

\begin{customthm}{4-1}[Compatibility Condition]\label{assumption4-1}
For some $\phi_0>0$ (independent of n) and for all $\eta$ satisfying $\|\eta_{S^c}\|_1 \leq 3\|\eta_{S}\|_1$, it holds that 

\begin{equation*}
\|\eta_{S}\|_1^2 \leq (\eta'(M_n)_S'W_n (M_n)_S\eta)s_{0}/\phi_0^2, 
\end{equation*}   
\end{customthm}

Assumption \hyperref[assumption4-1]{4-1} is referred as compatibility condition as in \cite{vandeGeer2014}. It also restricts the network topology (i.e complete network is still ruled out) but is weaker than assumption \ref{assumption3} + assumption \ref{assumption4}. Under assumption \hyperref[assumption4-1]{4-1}, we can still construct uniformly valid inference for de-bias estimator proposed in the next section even though consistent model selection is not guaranteed.

\subsection{Assumptions for Two Extensions}
\textbf{Heterogeneous Endogenous Effects Model with Cliques:}

The heterogeneous endogenous effects model with cliques can be written in matrix form as:
\begin{equation*}
D_n = \Big(M_n \circ D_n\Big)\eta_0 +  M_n D_n\gamma_0 + X_n\beta_0 + \epsilon_n
\end{equation*}

\begin{customthm}{1'}[Sparsity with Cliques]\label{assumption1'}
Among $n$ individuals in the network, let $S_n \subset \{1, 2, \cdots, n\}$ be the set of global leaders. Let $s_n = |S_n|$ be the number of elements in $S_n$. Assume:
\begin{equation*}
s_n = o\left(\frac{\sqrt{n}}{\log n}\right),\quad  \text{  as  } n \rightarrow \infty
\end{equation*}
\end{customthm}
Assumption \ref{assumption1'} relaxes the exact sparsity in assumption \ref{assumption1} without imposing any restriction on the number of local leaders. For example, it does not rule out situations where everyone is (locally) influential. Local leaders' influence will be captured by the $\gamma_0$, coefficient of classical spatial lag. The number of global leaders is restricted to be sparse and we can identify these leaders similar as in previous case. 

To ensure invertibility of the matrix $\Big(I_n-M_n \circ \eta_0 - M_n\gamma_0\Big)$, I modify assumption \ref{assumption2} as:
\begin{customthm}{2'}[SAR Restrictions with Cliques]\label{assumption2'}
In addition to assumption \ref{assumption2}, there exists an $\eta_{\max}<1$ such that $\|\eta_0+\gamma_0\|_\infty \leq \eta_{max}$
\end{customthm}

Similar to assumption \ref{assumption2}, this assumption excludes unit root processes. Since there exists a local level influence $\gamma_0$ in the network, global level influence $\eta_0$ need to be further bounded away from 1. As a result, equation (\ref{HEEMwHE}) can be transformed into the following:
\begin{equation}\label{13}
\begin{split}
E(D_n) =  X_n \beta_0 + \Big(M_n \circ X_n\Big)(\beta_0\eta_0) +  M_n X_n(\beta_0\gamma_0) +  \sum_{i  =2}^\infty\Big(M_n \circ \eta_0 +  \gamma_0 M_n \Big)^i\beta_0 X_n
\end{split}
\end{equation}

\begin{proposition}[First Stage Equivalence with Cliques]\label{proposition2}
Under assumption \hyperref[assumption1']{1'} and \hyperref[assumption2']{2'}  
\begin{equation*}
\begin{split}
E(D_n)  =  X_n\beta + (M_n \circ X_n)\tilde{\eta}^{*,\infty} + \sum_{i=1}^\infty\gamma_0^iM_n^iX_n\beta_0+ \sum_{i=1}^\infty M_n^i(M_n \circ X_n)\tilde{\tilde{\eta}}^{*,(i,\infty)},
\end{split}
\end{equation*}
where
\[\tilde{\eta}^{*,\infty}_j = \eta_{0,j} \tilde{g}_{k,j}^\infty(\eta_0, \gamma_0, \beta_0, X_n, M_n), \qquad \tilde{\tilde{\eta}}^{*,(i,\infty)}_j = \eta_{0,j} \tilde{h}_{k,j}^{(i,\infty)}(\eta_0, \gamma_0, \beta_0, X_n, M_n)\] 
for some function $\tilde{g}_{k,j}^\infty$ and $\tilde{h}_{k,j}^{(i,\infty)}$ depend on $\beta_0$, $\gamma_0$, $\eta_0$,  $X_n$, and $M_n$.
\end{proposition}

From proposition \ref{proposition2}, in order for the coefficient $\gamma_0$ introduced in equation (\ref{HEEMwHE}) to be identified. An extra instrument $M_nX_n$ need to be introduced. And thus we need to assume additional independence: 
\begin{customthm}{3'}[Independence with Cliques]\label{assumption3'}
$\big[W_n(M_n \circ X_n)_S, W_nM_nX_n\big]$ has full column rank.
\end{customthm}


To further incorporate exogenous and correlated effects, recall equation \eqref{HEEMwHEEC} can be written in matrix form as 
\begin{equation*}
D_n = \Big(M_n^* \circ D_n\Big)\eta_0 +  M_n^* D_n\gamma_{0}^{end} + M_n^* X_n\gamma_{0}^{exo} + X_n\beta_0 + \mu_c+ \epsilon_n
\end{equation*} 
Here, I define $M_n^*$ as the row sum normalized version of $M_n$. The main challenge for the identification is known as ``reflection problem" (\cite{Manski1993}). \cite{Bramoull2009} proposed the ``neighbor's neighbor" instruments as a solution. By taking the local difference, we can obtain the following form
\[(I - M_n^*)D_n = (I - M_n^*)\Big(M_n^* \circ D_n\Big)\eta_0 +  (I - M_n^*)M_n^* D_n\gamma_{0}^{end} + (I - M_n^*)M_n^* X_n\gamma_{0}^{exo} + (I - M_n^*)X_n\beta_0 + (I - M_n^*)\epsilon_n\]  
and invert the simultaneous equation, we have
\begin{equation*}
\begin{split}
E((I - M_n^*) D_n )  =  \Big(I - M_n^* \circ \eta_0 - M_n^* \gamma_{0}^{end} \Big)^{-1}\Big(\beta_0\cdot I + \gamma_{0}^{exo} M_n^* \Big)(I - M_n^*) X_n,
\end{split}
\end{equation*}

\begin{proposition}[First Stage Equivalence with Cliques + exogenous and correlated effects]\label{proposition3}
Under assumption \hyperref[assumption1']{1'} and \hyperref[assumption2']{2'} and assume $M_n^*$ is row sum normalized
\begin{equation*}
\begin{split}
E((I - M_n^*) D_n ) & = (I - M_n^*)X_n\beta_0 + (\gamma_{0}^{exo} +  \gamma_0^{end}\beta_0)M_n^*(I - M_n^*)X_n  + (M_n^* \circ X_n)\tilde{\eta}^{*2,\infty} \\
& + \sum_{i=1}^\infty(\gamma_0^{end})^i(\gamma_{0}^{exo}+\gamma_{0}^{end})M_n^{*(i+1)} (I - M_n^*)X_n\beta_0 + \sum_{i=1}^\infty M_n^{*i}(M_n \circ X_n)\tilde{\tilde{\eta}}^{*2,(i,\infty)}\\
\end{split}
\end{equation*}
where 
\[\tilde{\eta}^{*2,\infty}_j = \eta_{0,j} \tilde{g}_{2,k,j}^\infty(\eta_0, \gamma_{0}^{exo}, \gamma_{0}^{end}, \beta_0, X_n, M_n), \qquad \tilde{\tilde{\eta}}^{*2,(i,\infty)}_j = \eta_{0,j} \tilde{h}_{2,k,j}^{(i,\infty)}(\eta_0, \gamma_{0}^{exo}, \gamma_{0}^{end}, \beta_0, X_n, M_n)\] 
for some function $\tilde{g}_{2,k,j}^\infty$ and $\tilde{h}_{2,k,j}^{(i,\infty)}$ depend on $\beta_0$, $\gamma_{0}^{exo}$, $\gamma_{0}^{end}$,  $\eta_0$,  $X_n$, and $M_n$.\end{proposition}

\begin{customthm}{3'-1}\label{assumption3'-1}
Assume $\gamma_{0}^{exo} + \beta_0\gamma_{0}^{end} \neq 0$ and $\big[ W_n(M_n^* \circ X_n)_S, W_nM_n^*X_n, W_nM_n^{*2}X_n, W_nM_n^{*3}X_n\big]$ is full rank and $M^*$ is row sum normalizable.
\end{customthm}

Similar to proposition 4 in \cite{Bramoull2009}, assumption \hyperref[assumption3'-1]{3'-1} requires the independence between $M_n^*X_n$ and $M_n^{*2}X_n$ in order to use the ``neighbor's neighbors" as instruments to address the ``reflection" problem. Furthermore, the independence of $M_n^{*3}X_n$ will pin down the identification for correlated effects.

\textbf{Heterogeneous Endogenous Effects Model with Multiple Networks:}

The heterogeneous endogenous effects model with multiple networks can be represented in matrix form as follows:
\begin{equation*}
D_n = \sum_{j = 1}^q \left(M_n^j \circ D_n \right) \eta^j_0 + X_n\beta_0 +\epsilon_n
\end{equation*}
The number of coefficients in this model becomes $nq+k$ and the number of observed networks $q$ may also increase as the number of observations $n$ increases. As a result, the sparsity assumption will be imposed on both the influential individuals and the effective networks. I assume that some of the networks are completely irrelevant (i.e. $\eta_0^j = 0$) and that relevant networks are not necessarily passing influence for everyone (i.e. $\eta_0^j \neq 0$ but $\eta_{0,i}^j = 0$ for some $i$).

Second, to ensure invertibility, for any matrix norm $\|.\|$:
\begin{equation*}
\begin{split}
\left\| \sum_{j=1}^q\left(M_n^j \circ \eta^j_0 \right)\right\| \leq \sum_{j=1}^q \Big\|\left(M_n^j \circ \eta^j_0 \right)\Big\| \leq \sum_{j=1}^q \|\eta^j_0\|_\infty \Big\|\left(M_n^j \right)\Big\|
\end{split}
\end{equation*}
Because $M_n^j$ is the adjacency matrix such that each entry is 0 or 1, $\sum_{j=1}^q \|\eta^j_0\|_\infty<1$ guarantees the invertibility of $I  -\sum_{j=1}^q\left(M_n^j \circ \eta^j_0 \right)$.

\begin{proposition}[First Stage Equivalence -- Multiple Networks]\label{proposition4}
Under assumption assumption \hyperref[assumption1*]{1*} and assumption \hyperref[assumption2*]{2*} 
\begin{equation*}
\begin{split}
E(D_n) =  X_n \beta_{0} + \sum_{j = 1}^q \Big(M_n^j \circ X_n\Big)(\tilde{\eta}_0^{j}) 
\end{split}
\end{equation*}
where $\tilde{\eta}_k^j = \eta_k^j \tilde{g}^j(\eta_0, \beta_0, X_n, M_n)$ for some function $\tilde{g}^j$ depends on $\eta_0^j$, $\beta_0$, $X_n$, and $M_n^j$.
\end{proposition}

Third, I require $\left[X_n, \Big(M_n^1 \circ X_n\Big)_S, \Big(M_n^2 \circ X_n\Big)_S, \cdots, \Big(M_n^q \circ X_n\Big)_S\right]$ to be full rank. Compared with the standard model, this assumption requires the independence condition to hold across different networks. The four assumptions required for multiple networks are listed formally in the appendix as assumption assumption \hyperref[assumption1*]{1*}, \hyperref[assumption2*]{2*}, \hyperref[assumption3*]{3*}, and \hyperref[assumption4*]{4*}. 



\section{Estimators and Asymptotics}
\label{sec_estimator_2}
The proposed estimator is similar to the two-stage least square method but use LASSO in both stages. 


\begin{minipage}{\textwidth}
{\bf Two-Stage LASSO Estimator:}
\begin{itemize}
\item[-] First Stage:
\begin{equation} \label{firststage}
(\tilde{\beta},\tilde{\eta}) = \arg\min_{\beta, \eta}\|D_n - X_n \beta - \Big(M_n \circ X_n\Big)\eta\|^2_2+\lambda_1|\eta|_1
\end{equation}
\\\vspace{10pt}
Obtain a LASSO fitting $\hat{D}_n$
\begin{equation*}
\hat{D}_n = X_n \tilde{\beta} + \Big(M_n \circ X_n\Big)\tilde{\eta}
\end{equation*}
\item[-] Second Stage:
\begin{equation}\label{secondstage}
(\hat{\beta},\hat{\eta}) = \arg\min_{\beta, \eta} \|D_n - \Big(M_n \circ \hat{D}_n\Big)\eta - X_n\beta \|_2^2+\lambda|\eta|_1
\end{equation}
\end{itemize}
\vspace{5pt}
\end{minipage}
As shown in section \ref{sec_identification}, $\Big(M_n \circ D_n\Big)$ is correlated with $\epsilon_n$. Thus equation (\ref{HEEM}), equation (\ref{HEEMwHE}) and equation (\ref{HEEMwMN}) cannot be estimated directly using LASSO or sparse group LASSO. The instruments proposed in section \ref{sec_identification} are $[X_n, (M_n \circ X_n)_S]$. We do not observe the set $S$ but $[X_n, (M_n \circ X_n)]$ is a set of regressors that contains the valid instruments.  




\subsection{de-bias 2SLSS Estimator}
The estimator $\hat{\beta}$ and $\hat{\eta}$ suffer from LASSO shrinkage bias. Moreover, post model selection inference conditioning on the selected model $\hat{S}_n = \{i|\hat{\eta} \neq 0\}$ suffers from the omitted variable bias and thus is not uniformly valid \citep[see][]{LeebandPotscher2005, LeebandPotscher2008, leebandpotscher2009}. I construct a ``de-bias" estimator under my setting and derive the asymptotic distribution for it.  I propose the following de-bias LASSO estimator: 

\begin{minipage}{\textwidth}
\vspace{10pt}
{\bf de-bias 2SLSS Estimator:}
\begin{equation*}
\begin{split}
\hat{e} &= \hat{\eta} +\hat{\Theta}\tilde{\mathcal{X}}_n'(M_n \circ \hat{D}_n)'W_n(D_n - (M_n \circ D_n) \hat{\eta})/n \\
\\
\hat{b} &= \hat{\beta} - \frac{(X_n -  (M_n \circ D_n)\hat\gamma_\beta)'\Big((M_n \circ (\hat{D}_n - D_n)) \hat{\eta} \Big)}{(X_n -  (M_n \circ D_n)\hat\gamma_\beta)'X_n}
\end{split}
\end{equation*}
\vspace{5pt}
\end{minipage}
$\hat{\beta}$ and $\hat{\eta}$ are estimators from the 2SLSS. Define $W_n = \Big(I-X_n(X_n'X_n)^-X_n'\Big)$. $\tilde{\mathcal{X}}_n = \frac{1}{n}(M_n \circ D_n)'W_n(M_n \circ \hat{D}_n)$, $\hat{\Theta}$ are defined by the nodewise regression as in \cite{Meinshausen2006} on $\tilde{\mathcal{X}}_n$ and $\hat\gamma_\beta$ are again defined by the nodewise regression on between $X_n$ and $(M_n \circ D_n)$. Nodewise regression explores the correlation between the columns of the design matrix $\tilde{\mathcal{X}}_n$ by regressing each column on all the rest of the columns while penalizing the coefficients. An approximation of the inverse of the matrix $\frac{1}{n}\tilde{\mathcal{X}}_n'\tilde{\mathcal{X}}_n$ can be constructed based on nodewise regression. Further, define $\hat{S}_n = \{i|\hat{\eta} \neq 0\}$, which represents the LASSO selected active set. The estimators $(\hat{e}, \hat{b})$ are adjusted for the LASSO shrinkage bias and are a consistent estimator for $\eta$ and $\beta$. They are similar to the estimators proposed in \cite{vandeGeer2014}, but are constructed through a two-stage process. The de-bias estimator also differs from the two-stage estimators proposed in \cite{Zhu2016} due to spatial correlation.  


\begin{theorem}\label{theorem1}
Under \hyperref[assumption1]{assumption 1}, \hyperref[assumption2]{assumption 2}   and \hyperref[assumption4-1]{assumption 4-1}. There exist constant $c_1, c_2, c_3$ and $\iota:\|\iota\|_0 < \infty$ such that for the first stage tuning parameter $\lambda_1 \geq 2\sqrt{\frac{4\sigma^2c_1\Lambda_{min}^{-2}\log n}{n}}$ and second stage tuning parameter $\lambda \geq 2\sqrt{\frac{4\sigma^2c_3\Lambda_{min}^{-2}\log n}{n}} + 2c_2\lambda_1  + 2c_3\lambda_1$, the de-bias estimator  
\begin{equation*}
\begin{split}
&\sqrt{n}(\iota'\hat e - \iota'\eta_0) = \frac{1}{\sqrt{n}}\iota'\hat{\Theta}\tilde{\mathcal{X}}_n'\epsilon - \iota'\Delta \sim N(0, \sigma^2\iota'\hat{\Theta}\tilde{\mathcal{X}}_n'\Omega_n\tilde{\mathcal{X}}_n\hat{\Theta}'\iota)\\
& \sqrt{n}(\hat b - \beta_0) = \frac{(X_n -  (M_n \circ D_n)\hat\gamma_\beta)'\epsilon}{(X_n -  (M_n \circ D_n)\hat\gamma_\beta)'X_n} + \Delta_\beta \sim N\left(0, \sigma^2\frac{(X_n -  (M_n \circ D_n)\hat\gamma_\beta)'(X_n -  (M_n \circ D_n)\hat\gamma_\beta)}{\Big((X_n -  (M_n \circ D_n)\hat\gamma_\beta)'X_n\Big)^2}\right)
\end{split}
\end{equation*}
where $\|\Delta\|_\infty = o_p(1) $, $\|\Delta_{\beta}\|_\infty = o_p(1)$, $\tilde{\mathcal{X}}_n = \frac{1}{n}(M_n \circ D_n)'W_n(M_n \circ \hat{D}_n)$, $\Omega_n = \frac{1}{n}(M_n \circ \hat{D}_n)'W_n(M_n \circ \hat{D}_n)$, $\hat{\Theta}$ are defined by the nodewise regression on $\tilde{\mathcal{X}}_n$ and $\hat\gamma_\beta$ are defined by the nodewise regression on between $X_n$ and $(M_n \circ D_n)$.
\end{theorem}
Theorem \ref{theorem1} shows that the 2SLSS estimator achieves normality at the standard rate $\sqrt{n}$. The shifts $\Delta$ and $\Delta_\beta$ represent the bias from using nodewise regression and they are shown to be $o_p(1)$ with the proper choice of tuning parameters.   

\begin{theorem}\label{theorem2}
Under \hyperref[assumption1]{assumption 1-4}, there exist $\tilde\gamma>1$ and $\tilde\vartheta = \frac{1+\vartheta}{1-\vartheta}$. For $\lambda \geq 4(\tilde\gamma \vee \tilde \vartheta)\sqrt{\frac{\sigma^2c_3\Lambda_{min}^{-2}\log n}{n}} + 2c_2\lambda_1  + 2c_3\lambda_1$, 

\[\lim_{n \rightarrow \infty}\mathbb{P}(\hat{S}_n = S) = 1\]; 
\end{theorem}

\subsection{2SLSS for Two Extensions}

\begin{minipage}{\textwidth}
\vspace{10pt}
{\bf Two-Stage LASSO Estimator with Homogenous Effects:}
\begin{itemize}
\item[-] First Stage: for a pre-specified constant $k$,
\begin{equation*}
(\tilde{\beta},\tilde{\gamma}, \tilde{\eta}) = \arg \min_{\beta, \gamma, \eta}\left\|D_n - X_n \beta -  \sum_{i=1}^kM_n^i X_n\gamma^{1,i} - \sum_{i=0}^kM_n^i\Big(M_n \circ X_n\Big)\eta^{1,i}\right\|_2^2+\lambda\left(\sum_{i=1}^k|\eta^{1,i}|_1+|\gamma^{1,i}|\right)
\end{equation*}
\\\vspace{10pt}
Obtain a LASSO fitting $\hat{D}_n$
\begin{equation*}
\hat{D}_n = X_n \tilde{\beta} + M_n X_n\tilde{\gamma} + \Big(M_n \circ X_n\Big)\tilde{\eta}
\end{equation*}
\item[-] Second Stage:
\begin{equation*}
(\hat{\beta},\hat{\gamma}, \hat{\eta}) = \arg\min_{\beta, \gamma, \eta} \|D_n - M_n\hat{D}_n\gamma - \Big(M_n \circ \hat{D}_n\Big)\eta - X_n\beta \|_2^2+\lambda(|\eta|_1+|\gamma|)
\end{equation*}
\end{itemize}
\vspace{5pt}
\end{minipage}

From proposition \ref{proposition2}, the sparsity pattern can not be fully preserved by $(M_n \circ X_n)$ so additional structures like $M_n^i(M_n \circ X_n)$ need to be included in the first stage. Those terms represent the influence from global leaders but passing $i$th times through local leaders. By assumption \ref{assumption2'}, the influence represented by $M_n^i(M_n \circ X_n)$ decreases as $i$ increases. 

The second stage of 2SLSS with Cliques case can be viewed as a special case of the standard 2SLSS estimator. For example, one can rewrite the second stage as

\begin{equation*}
(\tilde{\beta},\tilde{\gamma}, \tilde{\eta}) = \arg \min_{\beta, \gamma, \eta}\left\|D_n - X_n \beta -  \left(\begin{array}{cc}
  M_n \hat D_n & \Big(M_n \circ \hat D_n\Big)
  \end{array}\right) \cdot \left(\begin{array}{c}
 \gamma\\
 \eta
 \end{array}\right)
\right\|_2^2+\lambda_1|\eta|_1 + \lambda_1|\gamma|_1
\end{equation*}

As a result, asymptotics as theorem \ref{theorem1} and theorem \ref{theorem2} follows.


%

To incorporate the structure of multiple networks, I propose the use of sparse group LASSO.


\begin{minipage}{\textwidth}
\vspace{10pt}
{\bf Two-Stage LASSO Estimator with Multiple Networks:}
\begin{itemize}
\item[-] First Stage:
\begin{equation*}
(\tilde{\beta},\tilde{\eta}) = \arg\min_{\beta, \eta} \left\{\left\|D_n - X_n\beta - \sum_{j = 1}^q (M_n^j \circ X_n) \eta^j   \right\|_2^2 + \left(\sum_{j = 1}^{q}\Big(\lambda_1 \|\eta^j\|_2 + \lambda_2 \|\eta^j\|_1\Big) \right)\right\}
\end{equation*}
\\\vspace{10pt}
Obtain a LASSO fitting $\hat{D}_n$
\begin{equation*}
\hat{D}_n = X_n \tilde{\beta} + \sum_{j = 1}^q (M_n^j \circ X_n) \tilde{\eta}^j 
\end{equation*}
\item[-] Second Stage:
\begin{equation*}
(\hat{\beta},\hat{\eta}) = \arg\min_{\beta, \eta} \left\{\left\|D_n - X_n\beta - \sum_{j = 1}^q (M_n^j \circ \hat{D}_n) \eta^j   \right\|_2^2 + \left(\sum_{j = 1}^{q}\Big(\lambda_1 \|\eta^j\|_2 + \lambda_2 \|\eta^j\|_1\Big) \right)\right\}
\end{equation*}
\end{itemize}
\vspace{5pt}
\end{minipage}

The sparse group LASSO introduces two tuning parameters, $\lambda_1$ and $\lambda_2$, to penalize both the $l_1$ and the $l_2$ norm in each network. Similar to the LASSO estimator, the geometric shape of the penalties allows the sparse group LASSO  to identify sparsity not only within each network (group) but also among networks (groups). In other words, some networks could be completely irrelevant (i.e. $\eta^j=0$) and within relevant networks, some individuals can have no influence on their neighbors (i.e. $\eta^j\neq 0$ but $\eta^j_i=0$ for some $i$).  

\begin{minipage}{\textwidth}
\vspace{10pt}
{\bf de-bias 2SLSS Estimator under Multiple Networks:}
\begin{equation*}
\begin{split}
\hat{e}^M &= \hat{\eta}^M+ \hat{\Theta}_Z\tilde{\mathcal{Z}}_n'\hat Z_n'W_n(D_n - Z_n \hat{\eta}^M)/n\\
\\
\hat{b}^M &= \hat{\beta}- \frac{(X_n -  Z_n\hat\gamma_\beta)'(\hat Z_n - Z_n \hat{\eta}^M )}{(X_n -  Z_n\hat\gamma_\beta)'X_n}
\end{split}
\end{equation*}
\vspace{5pt}
\end{minipage}

$\hat{\beta}$ and $\hat\eta_0^M = (\hat\eta^{1'}, \hat\eta^{2'}, \cdots, \hat\eta^{q'})'$ are estimators from the 2SLSS estimator with multiple networks. Let $\hat{Z}_n = \left[ (M_n^1 \circ  \hat{D}_n), (M_n^2 \circ \hat{D}_n), \cdots (M_n^q \circ  \hat{D}_n)\right]$, $Z_n = \left[ (M_n^1 \circ  D_n), (M_n^2 \circ D_n), \cdots (M_n^q \circ  D_n)\right]$. Define $\tilde{\mathcal{Z}}_n = \frac{1}{n}Z_n'W_n\hat Z_n$, $\hat{\Theta}_Z$ are defined by the nodewise regression as in \cite{Meinshausen2006} on $\tilde{\mathcal{Z}}_n$ and $\hat\gamma_\beta$ are again defined by the nodewise regression on between $X_n$ and $Z_n$.

\begin{theorem}\label{theorem3}
Under \hyperref[assumption1*]{assumption 1*}, \hyperref[assumption2*]{assumption 2*}   and \hyperref[assumption4-1*]{assumption 4-1*}. There exists constant $d_1, d_2, d_3$ and $\iota:\|\iota\|_0 < \infty$ such that for the first stage tuning parameter $\lambda_{1,1}  \geq 2\sqrt{\frac{4\sigma^2d_1\Lambda_{min}^{-2}(\log n +\log q)}{n}}$ and second stage tuning parameter $\lambda_{2,1}  \geq 2\sqrt{\frac{4\sigma^2d_3\Lambda_{min}^{-2}(\log n + \log q)}{n}} + 2d_2\lambda_{1,1}  + 2d_3\lambda_{1,1}$, the de-bias estimators  
\begin{equation*}
\begin{split}
\sqrt{n}(\iota'\hat e^M - \iota'\eta_0^M) = \frac{1}{\sqrt{n}}\iota'\hat{\Theta}_Z\tilde{\mathcal{Z}}_n'\epsilon - \iota'\Delta_M \sim N(0, \sigma^2\iota'\hat{\Theta}_Z\tilde{\mathcal{Z}}_n'\Omega_n^m\tilde{\mathcal{Z}}_n\hat{\Theta}_Z'\iota)
\end{split}
\end{equation*}
\begin{equation*}
\begin{split}
& \sqrt{n}(\hat b^M - \beta_0) = \frac{(X_n -  Z_n\hat\gamma_\beta)'\epsilon}{(X_n -  Z_n\hat\gamma_\beta)'X_n} + \Delta_{M,\beta} \sim N\left(0, \sigma^2\frac{(X_n -  Z_n\hat\gamma_\beta)'(X_n -  Z_n\hat\gamma_\beta)}{\Big((X_n -  Z_n\hat\gamma_\beta)'X_n\Big)^2}\right)
\end{split}
\end{equation*}

where $\|\Delta_M\|_\infty = o_p(1) $, $\|\Delta_{M,\beta}\|_\infty = o_p(1)$ and $\tilde{\mathcal{Z}}_n = \frac{1}{n}Z_n'W_n\hat Z_n$ and $\Omega_n^m = \frac{1}{n}\hat{Z}_n'W_n\hat Z_n$. $\hat{\Theta}_Z$ are defined by the nodewise regression on $\tilde{\mathcal{Z}}_n$ and $\hat\gamma_\beta$ are defined by the nodewise regression on between $X_n$ and $Z_n$.
\end{theorem}
Theorem \ref{theorem3} requires the rate of convergence for sparse group LASSO at the first stage. This is proved in Lemma \ref{FiniteSampleBoundFirstMultiple} in the appendix. In group LASSO, $\lambda_{2,1}$ controls the convergence of the $l2$-norm and thus it might be tempting to set it larger than $\lambda_{1,1}$ to achieve $l_2$ convergence. However, group LASSO requires the number of regressors in each group to be finite. In my estimator, the number of regressors in each group equals to $n$. In the next theorem, I show that $\lambda_{2,1}$ will need to be chosen in the same order as $\lambda_{1,1}$ in order to achieve consistent selection.  

\begin{theorem}\label{theorem4}
Let $\tilde c = \frac{\lambda_{2,1}}{\lambda_{2,2}}$ and $\tilde\vartheta^{mul}_1 = \frac{1+\vartheta^{mul}_1}{1-\vartheta^{mul}_1}$. Choose $\lambda_{2,1}$ and $\lambda_{2,2}$ such that $\tilde c> \frac{\vartheta^{mul}_1}{1-\vartheta^{mul}_1}$. Under \hyperref[assumption1*]{assumption 1*-4*} and for $\lambda_{2,1} \geq 4(\tilde\gamma \vee \tilde \vartheta^{mul}_1)\sqrt{\frac{\sigma^2d_3\Lambda_{min}^{-2}(\log n + \log q)}{n}} + 2d_2\lambda_{1,1}  + 2d_3\lambda_{1,1}$, there exists $\tilde\gamma>1$,
\[\lim_{n \rightarrow \infty}\mathbb{P}(\hat{S}_n = S) = 1\] 
\end{theorem}
It is worth pointing out that assumption \ref{assumption4*} is a weaker assumption than assume the full irrepresentable condition on the design matrix 
\begin{equation*}
\Sigma^{mul} := \frac{1}{n}[M_n^1, M_n^2, \cdots, M_n^q]' W_n[M_n^1, M_n^2, \cdots, M_n^q] 
\end{equation*}
As the number of regressors in each group also goes to infinity as the number of groups goes to infinity, the multicollinearity in $\Sigma^{mul} $ can be severe. Instead, assumption \ref{assumption4*} decomposes the multicollinearity into between-group and within-group and thus allows a more flexible dependence structure. As a result, sparse group LASSO can be used to recover the influential individuals as well as the effective networks that generate spillovers.

\section{Simulations}
\label{sec_simulation_2}
In this section, I report Monte Carlo simulation results for the models proposed above. My results are robust when applied to networks generated by different algorithms  and to networks of different sizes.

\subsection{Heterogeneous Endogenous Effects Model}
First, I use the Erdos-Renyi algorithm to simulate a network of size $n$. Individuals are added into the graph one at a time. When one individual is added to the network, she has probability $p$ of generating a link with all existing individuals independently. I choose $p=0.1$ and $p=0.2$ in the simulation. I avoid a large $p$ because collinearity among regressors may arise when links become very dense, violating assumption \ref{assumption3} or \ref{assumption4}. 

I set the first 5 individuals to be influential by letting their coefficients $\eta_j$ be non-zero. To guarantee the existence of endogenous effects, I arbitrarily specify the connections among these five individuals.  The adjacency matrix $M_n$ for the five influential individuals is given in the appendix. If the connections among these five individuals are not fixed, there is a possibility that no connections are formed among these five and thus there is no endogeneity in the network. The results will be too good in such a case. The true parameters are fixed as $\beta_0 = 3$, $\eta_{0,1} = \eta_{0,2} = \eta_{0,3} = \eta_{0,4} = \eta_{0,5} = 0.5$, and $\eta_{0,j}=0$ for $j>5$. Individual characteristics $X_n$ are generated from a standard normal distribution. Individual outcomes $Y_n$ are then generated as $Y_n = (I-M_n\circ\eta_0)^{-1}(X_n\beta_0+ \epsilon_n)$  where $\epsilon_n$ is drawn independently from a standard normal distribution. 

I use $(M_n, X_n, Y_n)$ as observations and apply my two-stage LASSO estimator. I construct the de-bias 2SLSS estimator and repeat the above process 200 times in a manner similar to \cite{vandeGeer2014}. I report the average coverage probability (Avgcov) and average length (Avglength) of confidence intervals for the coefficients for influential individuals, $\{\eta_1, \cdots, \eta_5\}$, the coefficient for individual characteristics, $\beta_0$, and the coefficients for non-influential individuals, the $\eta_j$s ($j>5$). For example:
\begin{equation} \label{avgcov}
\text{Avgcov  }S_0 = s_0^{-1} \sum_{j \in S_0} \mathbb{P}[\eta_{0,j} \in CI_j]
\end{equation}
\begin{equation}\label{aß†vglength}
\text{Avglength  } S_0 = s_0^{-1} \sum_{j \in S_0} length( CI_j)
\end{equation}

I separately report the average coverage and average length for each of the five influential individuals. As shown in appendix table A1, the coverage is around the nominal $95\%$ level and the length of the confidence intervals decreases as the sample size grows. 

Since we can construct confidence intervals for all $n$ coefficients, joint inference can be performed under the control of False Discover Rate (FDR). As shown in equation (\ref{power}), the power reported in appendix table A1 represents the average percentage in the active set (i.e. $\{1,2,3,4,5\}$) that is significant after controlling for the False Discover Rate (FDR) at 5\% using the Benjamini-Hochberg method. The FDR reported in appendix table A1 represents the average percentage of the non-active set (i.e. $\{6,7,\cdots, n\}$) that is  significant after controlling the FDR at 5\% using the Benjamini-Hochberg method. The exact definition is as in equation (\ref{fdr}).

\begin{equation}\label{power}
\text{Power } = s_0^{-1} \sum_{j \in S_0} \mathbb{P}[H_{0,j} \text{ is rejected}]
\end{equation}
\begin{equation}\label{fdr}
\text{FDR  } =  \sum_{j \in S_0^c} \mathbb{P}[H_{0,j} \text{ is rejected}]/\sum_{j=1}^n \mathbb{P}[H_{0,j} \text{ is rejected}]
\end{equation}

The power varies because the networks change when the sample size increases. It is strictly increasing when the network is sparse (i.e. $p=0.1$). The power decreases in the $p=0.2$ case as the problem of endogeneity increases when the network is dense. The empirical FDR is controlled well, which all under the $5\%$ rate. Notice that the confidence interval's length is large when the sample size equals 50. This is because when the number of individuals is small, some individuals might only connect to 1 or 2 other individuals. This means that the regressors that represent this individual are all 0s except for a small numbers of non-zero terms, which leads to a large standard error.

The two-stage LASSO estimator requires the choice of two tuning parameters (i.e. the two $\lambda$s from both stages as in section 4.1). Moreover, when calculating $\hat{\Theta}$ in the de-bias 2SLSS estimator (section 4.2) and using the nodewise regression, one also need to choose a tuning parameter. I use a benchmark choice of $\lambda_{nodewise}\gtrsim \sqrt{\log(n)/n} $ and $\lambda_1\gtrsim \sqrt{\log(n)/n}$ in the nodewise regression and first stage. Then I use cross-validation to pick $\lambda$ in the second stage.

I further increase the number of influential individuals to 10 and report the results in appendix table A2. Again, to guarantee the existence of endogeneity, the adjacency matrix for these ten individuals is set as shown in the appendix. All average coverages and average confidence interval lengths are separately reported for these ten individuals. The choice of the tuning parameters is similar to those used to generate appendix table A1 for networks with 50 and 200 individuals. For networks with 500 individuals, I use benchmark $\lambda$ to replace cross validation in the second stage. 

As shown in appendix table A2, all coverages are very close to the nominal levels. The average lengths of confidence intervals is slightly larger compared with appendix table A1. This is due to the increase in influential individuals; it is more difficult to differentiate them from those irrelevant individuals.

Appendix table A3 presents the result when a network is generated using the Watts-Strogatz mechanism or the ``small world" network. Define the $pN$ (even number) as the mean degree for each node and a special parameter $\omega = 0.4$. The Watts–Strogatz mechanism works as follows:
\begin{itemize}
\item[-] construct a graph with N nodes each connected to $pN$ neighbors, which $\frac{pN}{2}$ on each side. 
\item[-] For each node $n_i$, take every edge $(n_i, n_j)$ with $i<j$ and rewrite it with probability $\omega$. Rewrite means replace $(n_i, n_j)$ with $(n_i, n_k)$ where $k$ is choosing uniformly among all nodes that are not currently connected with $n_i$ 
\end{itemize}
The influential individuals are chosen as the 1st, 5th, 15th, 40th and 50th individuals in the network. As shown in appendix table A3, my estimator is robust under a ``small world" algorithm. Nominal level is reached as the size of the network grows and the length of confidence intervals is slightly smaller than in the standard case.

\subsection{Heterogeneous Endogenous Effects Model with Cliques}
Appendix table A5 presents results for the heterogeneous endogenous effects model with cliques. The outcome variable $Y_n$ is now generated as  $Y_n = (I-M_n\circ\eta_0 - M_n\gamma_0)^{-1}(X_n\beta_0+ \epsilon_n)$. The coefficient of the homogeneity effects $\gamma_0$ is set at 0.05. 

The choice of the tuning parameters is similar to that used to generate appendix table A1 for networks with 50 and 200 individuals. For networks with 500 individuals, I use benchmark $\lambda$ (i.e. $\lambda \gtrsim \sqrt{\log(n)/n}$) to replace cross validation in the second stage. 

The coverage is above the $95\%$ nominal level in all cases. I also report the mean coverage and average length of the confidence å†interval for the coefficient of the homogeneous effects. My model gives above $95\%$ coverage in all cases. I also report the empirical probability of rejecting a null hypothesis of zeros effects at $95\%$ nominal level. The probability of rejecting the test converges to 1 when the sample size grows to 500.    

\subsection{Heterogeneous Endogenous Effects Model with Multiple Networks}
In this Monte Carlo exercise, I include two different networks generated by the Erdos-Renyi algorithm, where one is influential and the other is not. I use the two-stage LASSO estimator with multiple networks to estimate the parameters. The sparse group LASSO requires two tuning parameters, one for the $l_2$ norm and the other for the $l_1$ norm. I set the two parameters to be equal to each other as the correlations among the columns of the adjacency matrices are very small. The choice of tuning parameters is similar to that used to generate appendix table A1 for networks with 50 and 200 individuals. For networks with 500 individuals, I use benchmark $\lambda$ instead of cross-validation in the second stage. Appendix table A4 summarizes the results. As in previous results, all coverages exceed the nominal 95\% level. 

I report the empirical probabilities such that at least one individual is detected in a given network controlling for the FDR at 5\% using the Benjamini-Hochberg method. I also report the average number of detections conditioning on at least one individual who is detected in a given network. Appendix table A4 shows that network 1, which is the relevant network, is more likely to be detected in all cases than network 2, the irrelevant network. The average number of identified individuals for network 1 is also more than that of network 2.

\section{Empirical Application}
\label{sec_application}
I use the proposed estimator to study the importance of different networks in spreading the participation in a micro finance program within rural Indian villages.  I show that different kinds of networks have different effects on individuals decisions. I identify the influential individuals in each village. My analysis shows that leaders among agricultural laborers, Anganavadi teachers, construction workers, small business owners and mechanics are very likely to be influential in the villages.

\subsection{Background}

A non-profit organization named Bharatha Swamukti Samsthe (BSS) has been running micro finance programs in rural southern Karnataka, India since 2007. It provides small loan products to poor women and, through them, to their families. The villages covered by the program are geographically isolated and heterogeneous in terms of caste. 

When BSS initially introduces a micro finance program to a village, the credit officers of BSS first approached a number of ``predefined leaders", such as teachers, shopkeepers and village elders. BSS held a private meeting with these leaders and explained the program. Then these predefined leaders passed the information onto other villagers. Those who were interested in the program and contacted BSS were trained and assigned to groups to receive credit. Each group consisted of 5 borrowers and group members were jointly liable for loans. Loans were around 10,000 rupees (approximately \$200) at an annualized rate of approximately 28\%. Note that 74.5 percent of the households in rural area said the monthly income of their highest earning member is less than 5,000 rupees (source: Socio-Economic Caste Census-2011). This loan had to be repaid within 50 weeks. 

In 2006, 75 villages in Karnataka were surveyed 6 months before the initiation of the BSS micro finance program. This survey consisted of a village questionnaire and a detailed follow-up survey conducted among a subsample of villagers. The village questionnaire gathered demographic information on all households in a village including GPS coordinates, age, gender, number of rooms, whether the house had electricity, and whether the house had a latrine. The data set also contains information on the ``pre-defined leaders" set who helped spread the information to the entire village. The follow-up survey collected data from a villager sample stratified according to age, education level, caste, occupancy, etc. It also asked questions about social network structures along 12 dimensions, including:

\begin{itemize}
\item[-] Friends: Name the 4 non-relatives whom you speak to the most.  
\item[-] Visit-go: In your free time, whose house do you visit?  
\item[-] Visit-come: Who visits your house in his or her free time? 
\item[-] Borrow-kerorice: If you needed to borrow kerosene or rice, to whom would you go? 
\item[-] Lend-kerorice: Who would come to you if he/she needed to borrow kerosene or rice? 
\item[-] Borrow-money: If you suddenly needed to borrow Rs. 50 for a day, whom would you ask? 
\item[-] Lend-money: Who do you trust enough that if he/she needed to borrow Rs. 50 for a day you would lend it to him/her? 
\item[-] Advice-come: Who comes to you for advice? 
\item[-] Advice-go: If you had to make a difficult personal decision, whom would you ask for advice? 
\item[-] Medical-help: If you had a medical emergency and were alone at home whom would you ask for help in getting to a hospital? 
\item[-] Relatives: Name any close relatives, aside from those in this household, who also live in this village. 
\item[-] Temple-company: Do you visit a temple/mosque/church? Do you go with anyone else? What are the names of these people?
\end{itemize}

\begin{figure}
\centering
\includegraphics[scale=0.4]{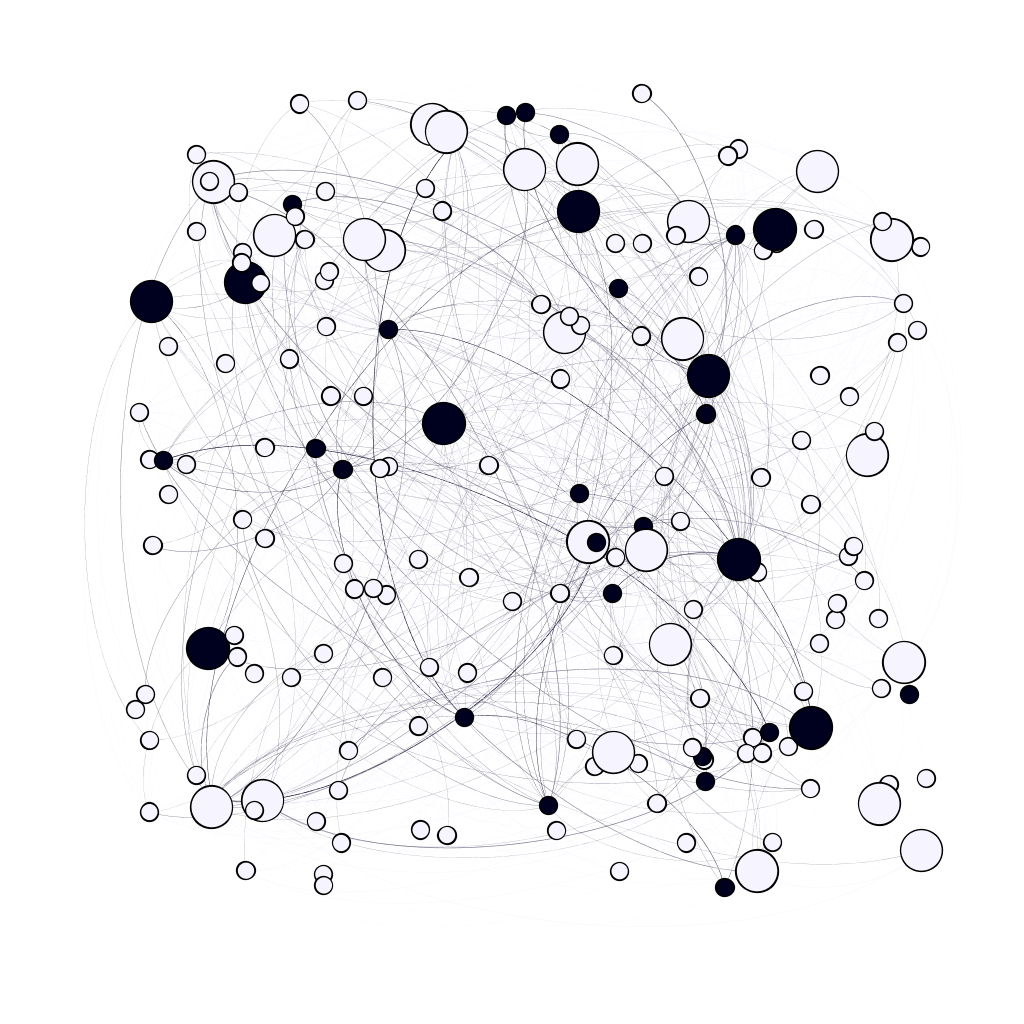}
\caption{Network in Village 1}
    \label{fig:village1}
\end{figure}

For the 43 villages where micro finance was introduced by the time of 2011, BSS also collects information on which villagers have joined the program. These survey questions reveal the underlying structures for connections among any two individuals in the network. Figure \ref{fig:village1} presents all those connections at the household-level in a graph. Each node in the graph represents a household. A black node indicates that the household joined the micro finance program, while a white node indicates that it did not.  Bigger nodes represent those households in which at least one family member has been chosen as being among the ``pre-defined leaders". An edge between two nodes signifies that the two nodes are connected in at least one of the 12 networks. The darker the color of the edge, the more connections it represents. 

This dataset provides an ideal framework for application of the heterogeneous endogenous effects model. First, it allows me to model endogenous effects. An individual may decide to join the micro finance program if her neighbors or friends plan to join. Second, the endogenous effects are individual specific. Given the diversity of the villagers, it is possible that some villagers are more influential than others. Third, it allows me to implement the heterogeneous endogenous effects model with multiple networks. The questions asked regarding multiple dimensions of the network structure allow me to explore which network is most influential.  

\subsection{Data}
In this empirical study, I focus on the 38 villages that have been introduced to the micro finance programs by BSS. For each village, I can observe both its social network structure and the villagers' decisions about joining the program. I drop the data for one village (Village 46) that contains incorrect entries on the index of households. Appendix table \ref{apdxtab12} summarizes the descriptive statistics for each village.

Among the 12 questions about the social network structure, 4 pairs essentially capture the same connections among the villagers \footnote{Assuming every villager truthfully answers a pair of questions, the adjacency matrices associated with  each question are the same. It is also plausible to treat villagers' answers to each question as a separate directed graph. However, these questions do not allow for clear determination of the directions. For example, if villager $A$ visits villager $B$'s house, it is not clear whether villager $A$ influences villager $B$ or vice versa}. Therefore, I consolidate each pair of questions into one dimension:

\begin{equation*}
\begin{split}
\text{Visit-go-come} \quad & \begin{cases} \text{In your free time, whose house do you visit? }\\ \text{Who visits your house in his or her free time? }\end{cases}\\
\text{Borrow-Lend-kerorice} \quad & \begin{cases} \text{If you needed to borrow kerosene or rice, to whom would you go? }\\ \text{Who would come to you if he/she needed to borrow kerosene or rice?}\end{cases}\\
\text{Borrow-Lend-money} \quad &\begin{cases} \text{If you needed to borrow Rs.50 for a day, whom would you ask? }\\ \text{Who do you trust enough that if he/she needed to borrow Rs.50 for a }\\ \qquad \text{day  you would lend it to him/her?}\end{cases}\\
\text{Help decision} \quad &\begin{cases} \text{Who comes to you for advice? }\\ \text{If you had to make a difficult personal decision, whom would you ask  } \\ \qquad \text{for advice?}\end{cases}
\end{split}
\end{equation*}


I restructure all the data at the household level as only women are allowed to apply for the micro finance program because the goal of BSS is to support families through the women in them. As a result, a woman's decision to join or not join the micro finance program becomes her family's decision. A connection between two villagers becomes a connection between two families. A ``predefined leader" is a villager selected by BSS to help spread information about the micro finance program to the other villagers. At the household level, I use the term ``predefined leader" for a household that contains at least one such villager.     

\subsection{Sparsity and Equilibrium}
To demonstrate how my method identifies influential households, I model families' decisions regarding joining the micro finance program as a network game with Bayesian Nash Equilibrium. For household $i$, let $d_i^*$ be the expected probability that $i$ chooses to join the micro finance program. The decision of household $i$ depends on its neighbors' decisions as well as the types of connections between them. The decision also depends on its characteristics $X_i$ and on unobserved information $\epsilon_i$. Formally, it can be written as:
\begin{equation*}
d_i^* = \sum_{l \in N_i}d_l^*(\sum_{j = 1}^q \eta_l^j)+ x_i\beta + \epsilon_i
\end{equation*}
Rewritten in matrix form:
\begin{equation*}
D_n^* = \sum_{j = 1}^q \left(M_n^j \circ D_n^* \right) \eta^j + X_n\beta +\epsilon_n
\end{equation*}

I assume that only a small number of households are influential over their neighbors. Leaders and followers are usually observed in those rural villages. Big decisions are often made by the village elders or by the more educated among the villagers. BSS recognized the importance of leaders and gathered a group of predefined leaders, asking them to inform the rest of the villagers about their program. I do not consider the local level influence in these villages given the size and how complicated the network structures are. Households are closely connected by these 8 networks as shown in Figure \ref{fig:village1} and there is no form of clique visible.         

Because the villages are considered geographically isolated, I apply my estimator separately to each of the 38 villages. I use the number of rooms per person in a household as the independent variable $X_n$. Number of rooms per person is a proxy for the wealth in the family. As shown in table \ref{table6}, it is negatively correlated with the decision to join the micro finance program. The richer the family, the less likely the family is to participate in the micro-finance program. On the other hand, it is arguably an exogenous variable to other factors (career, education, etc) that might affect the decision to join the micro-finance program.  I further check the robustness of my independent variable by including additional controls. The adjacency matrix $M_n^j$ is constructed from the questions in the survey. Households $i$ and $k$ are connected in network $j$ if either $i$ or $k$ reported the other in question $j$. Finally, $d_i^*$ is replaced with the household's decision.

\begin{table}[htbp] \caption{Predictive Power of Characteristics $X_n$} 
\begin{center}
\scalebox{0.8}{\label{table6}
\hspace{-1cm}
\begin{tabular}{lcccc} 
\hline \hline

     &&  (1) &  (2)     &   (3) \\
     &&  Participate& Participate & Participate\\
\midrule
Average Num. rooms x100  &&  -8.19$^{***}$   &  -7.12$^{***}$  &  -3.36$^{***}$ \\
  &&  (1.30)  & (1.30)& (1.12)\\
Household Size x100     &&      & 0.45$^{**}$  & 0.40$^{**}$\\ 
 &&    & (0.21)& (0.20)\\
Electricity x100    &&     &   &1.29\\ 
 &&    & & (1.26)\\
Latrine x100    &&      &   &-5.66$^{***}$\\ 
 &&    & & (1.38)\\
Average Num. workers x100   &&      &   &0.64$^{*}$\\ 
 &&    & & (0.50)\\
Average age x100    &&      &   & -0.27$^{***}$\\ 
 &&   & & (0.03)\\
Village Fixed Effects&&   Y   & Y  & Y\\
\midrule
$n$  &&    8,375   &  8,375  &  8,375\\
\# of villages  &&    37   &  37  &  37\\
$R^2$  &&    0.05   &  0.05  & 0.06 \\

\hline\hline
\end{tabular}
}
\end{center}
\scriptsize{ \justify Table \ref{table6} provides a robustness check for variable $X_n$: Average Num. rooms, which used to construct instruments. Standard errors in parentheses * $p<0.1$, ** $p<0.05$, *** $p<0.01$. Standard deviation clustered at village level. Dependent variable is households' decision on whether to join the micro finance program or not. All design control village fixed effects.}
\end{table}

The instruments are constructed as $\left(M_n^j \circ X_n \right)$ for $j = 1, 2, \cdots, 8$. I use the heterogeneous endogenous effects model with multiple networks to: 1) Identify the effective networks affecting a household's decision and 2) Identify that households that are leaders in the village and study the association between observable characteristics and leader status. If a new program is going to try to recruit these households, the organizers can target those influential households and try to persuade them to join first.     

\subsection{Results}
\subsubsection{Identifying Effective Networks}
First, I study how LASSO selects networks. I define a coefficient for a household's endogenous effect in a network as significant according to two different criteria. The first criterion, ``Cross-Validation", determines a coefficient to be significant if LASSO predicts the coefficient to be non-zero after cross-validation. The second criterion, ``de-bias", first constructs a bias-adjusted coefficient and calculates its standard error. It then determines a coefficient to be significant if the Benjamini-Hochberg method rejects the null hypothesis of zero effect at the 5\% false discovery rate. A network is defined as significant if at least one coefficient for heterogeneous endogenous effects in this network is significant.  

Table \ref{table7} presents the empirical probability of the 8 networks being significant among the 37 villages. Note that certain types of networks (such as visit go-come) are more likely to pass influence then others (such as temple company). For example, by cross-validation criterion, the visit go-come network is detected as significant in 26 out of the 37 villages (i.e. 70\%) whereas temple company is detected as significant in none of the 37 villages (i.e. 0\%). I also present the average number of households associated with significant endogenous effects in each significant network. For example, according to the cross-validation criterion, an average of 101 households are detected having significant coefficients associated with the visit go-come network among the 26 village. The average of the absolute magnitude of the heterogeneous endogenous effects is also reported. 

In terms of variable selection, if Assumption 4 holds, the cross-validation criterion may consistently select the truly influential households with high probability even in a finite sample. On the other hand, the de-bias criterion is likely to be conservative because of its use of the false discovery control process. In terms of coefficients estimated, de-bias estimators are asymptotically consistent. On the other hand, estimates based on the LASSO estimator suffer from shrinkage bias and are not consistent.

\begin{table}[htbp]\caption{Influential Networks} 
\begin{center}
\scalebox{0.76}{\label{table7}
\begin{tabular}{l*{11}{c}} 
\hline \hline

     &&  &visit & friendship   &   borrow-lend   & borrow-lend     &  relatives &   help     &  medical & temple   \\
     &&   &go-come &             &   keroric    &  money        &      & decision &  help & company  \\
\midrule
Cross \footnotemark[1] & probability \footnotemark[3]   &&   70\% &   54\%   &   38\%  &  49\%   &  22\%   & 30\%   & 14\% & 0\%     \\
 Validation &  identified   \footnotemark[4]        &&   101  &   89   &  88  &  86  &  69   & 70   &   80  & 0 \\
 \midrule
   \multirow{3}{*}{De-bias   \footnotemark[2]}  & probability \footnotemark[3]  &&   54\% &   35\%  &   30\%  &  19\%  &   16\% & 8\%  & 5\% & 0\%      \\
   & identified \footnotemark[4] &&   11  &  8   & 8   & 9  & 7  &  14  &  8  &  0  \\
   & magnitudes   \footnotemark[5]  &&  0.0166 & 0.0095 & 0.0070 & 0.0061 & 0.0050 & 0.0040 & 0.0028 & 0.0018    \\

\hline \hline

\end{tabular}
}
\end{center}
\scriptsize{\justify Table \ref{table7} reports the probability of detection for different networks among the 38 villages. A network is detected as influential if at least one leader is detected within this network. 1. Cross Validation represents those networks detected by lasso using cross validation. 2. De-bias represents those networks detected by significant de-bias estimators under FDR control. 3. Probability reports the empirical probability that at least one coefficient $\hat{e}_i^j$ is significant in network $j$. 4. Identified reports the averaged number of significant $\hat{e}_i^j$ in the network $j$ conditioning on the network being detected. 5. Magnitudes is the mean of $|\hat{e}_i^j|$ and represents the average endogenous effects through network $j$} 
\end{table}

The results in Table \ref{table7} suggest villagers are more likely to discuss the micro-finance program when they visit each other, chat with friends, and meet with people to whom they are economically connected. Villagers are not likely to talk about the micro finance program when they go to the temple or need medical help.

\subsubsection{Identifying Influential Households}

Second, I focus on how LASSO selects households. I compare the LASSO selected influential households with the BSS selected ``predefined leaders". It is important to point out that these ``predefined leaders" are \emph{not} necessarily influential villagers in a network.  Recall that predefined leaders are a set of villagers that BSS select to help spread the information about the micro finance program. The fact that a villager is selected as a ``predefined leader" to \emph{pass information} about the micro finance program does not a priori guarantee her or her family's \emph{influence} -- her decision to join the micro finance program may not lead to her neighbors' decisions to join. In the analyses below, I will examine how influential villagers are associated with ``predefined leaders" and explore their potential differences. 

{\bf 1. Influential Predefined Households}\\
In table \ref{table:leader}, I report results indicating that influential households selected by LASSO partly overlap with ``predefined leaders". This is intuitive because some ``predefined leaders" such as school headmasters and village elders are highly respected figures in a village. Therefore, their decisions are likely to be followed by others in the village. On average, BSS selected 27 villagers as ``predefined leaders" in each village. In comparison, Cross-Validation criterion selects around 136 villagers and de-bias criterion selects around 19. Furthermore, on average, 19 out of 136 influential villagers (i.e. 14\%) selected by Cross-Validation criterion are also BSS ``predefined leaders"; 3 out of 19 influential villagers (i.e. 17\%) selected by de-bias criterion are also BSS ``predefined leaders". The likelihood of selected by the two methods are both higher than the percentage of predefined leaders in the entire village (11\%). Comparing with a random guess of influential individuals, table \ref{table:leader} suggests the LASSO detected influential individuals are more likely to overlap with the predefined leaders. In Table \ref{table11} below, I show that small business owners are more likely to be both influential and selected as ``predefined leaders".

\begin{table}[htbp]
\caption{Coverage of predefined leaders}
\begin{center} 
\scalebox{0.8}{\label{table:leader}
\begin{tabular}{l*{4}{c}} 
\hline \hline
    && \multicolumn{2}{c}{\% of predefined leaders among:} & \multirow{2}{*}{Average number of discovery \footnotemark[3]}\\
    \cline{3-4}
    &&  LASSO detected \footnotemark[1] & entire village \footnotemark[2] &    \\
\midrule

 Cross Validation \footnotemark[4]    &&   14\% & 11\%  & 136    \\
 De-bias   \footnotemark[5]  &&   17\% &  11\% &  19   \\
\hline \hline
\end{tabular}
}
\end{center}
\scriptsize{\justify Table \ref{table:leader} depicts the overlapping between influential households selected by LASSO and ``predefined leaders". Predefined leaders are a set of villagers defined by BSS, who helped spread the information about the micro-finance program. 1. LASSO detected reports the percentage of households detected by LASSO and also selected as ``predefined leaders" in total LASSO detected households. 2. Entire village reports the percentage of ``predefined leaders" among the entire village. 3. Average number of discovery reports the total number of individuals discovered by LASSO using each method. 4. Cross Validation represents those individuals identified from lasso using cross validation. 5. De-bias represents those individuals identified from significant de-biased estimators controlling FDR. 6. The average number of predefined leaders in one village is 27.}
\end{table}

{\bf 2. Influential Non-Predefined Households}\\
In this and the following section, I focus on understanding the differences between the influential households selected by LASSO and the ``predefined households" selected by BSS. I investigate the likelihood that a household being selected by LASSO or by BSS, as associated with the careers of its family members. Table \ref{table9} and \ref{table10} present linear regression results using career dummy variables of family members to explain whether a household is selected as ``predefined leader" (Column 1 in table \ref{table9}), whether a household is selected by LASSO as influential using cross-validation (Column 2 in table \ref{table9}), and whether a household is selected by LASSO as influential using de-bias estimator (Column 3 in table \ref{table9}). The full results of these regressions are reported in appendix table A7.


Table \ref{table9} summarizes all careers that have a significant impact ($p<0.001$) on the likelihood of a household being selected by LASSO as influential using de-bias method. Agriculture is the backbone of those rural villages and over 67\% of the villagers are agricultural laborers. The prior that there exists influential individuals among agricultural laborers coincides with the LASSO detection. Anganwadi teacher is a set of groups that provides pre-school education to the children. They are part of the government's health care system in the rural areas. Note that except for small business owners and Anganavadi teacher all the other careers in this table are either not significantly or negatively associated with the likelihood of a household being selected as ``predefined leaders". Compare with Anganavadi teacher and small business owners, regular teachers and animal skin business owners are not targeted by BSS as ``predefined leaders" but LASSO estimators indicate they may also become influential. Other careers that are correlated with LASSO selection include construction/mud worker, truck/tractor driver, factory worker, daily labor and wood cutter. Industrial workers are likely to form circles of collaboration due to the natural of their jobs and thus it seems compelling that they are selected as influential individuals.    

Table \ref{table10} summarizes all careers that have a significant impact ($p<0.01$) on the likelihood of a household being selected by BSS as being among the ``predefined leaders". Poojari are Indian priests in those villages and they are very likely to be included as ``predefined leaders". However, they are not likely to influence people to join the micro finance program. Other careers as tailor, hotel workers, veteran and barber are included as ``predefined leaders" because individuals doing these jobs can spread information quickly in the village. However, LASSO does not find these individuals to be influential.    

Table \ref{table11} reports the counter factual study when selected leaders all decide to join the micro-finance program. The participation rate for non-leaders in the data is 16\%. When all ``predefined leaders" decide to join, the participation rate for non-leaders will increase to 20\%. And when all LASSO selected leaders decide to join, the participation rate for non-leaders will further increase to 33\%.            




\begin{table}[htbp] \caption{Second Stage: LASSO selected leaders' careers} 
\begin{center}
\scalebox{0.8}{\label{table9}
\begin{tabular}{l*{4}{c}} 
\hline \hline
     &&  Predefined  & \multicolumn{2}{c}{ Selected by LASSO }\\
     \cline{4-5}
     && leaders      &  Cross-validation & De-bias \\
\midrule
Agriculture labour                  &&0.00&0.31$^{***}$&0.06$^{***}$\\
&&(0.01)&(0.01)&(0.00)\\
Anganavadi Teacher                  &&0.14$^{*}$&0.11&0.13$^{***}$\\
&&(0.06)&(0.07)&(0.04)\\
Construction/mud worker               &&0.00&0.17$^{***}$&0.15$^{***}$\\
&&(0.02)&(0.03)&(0.02)\\
Truck/Tractor Driver                &&-0.03&0.16$^{***}$&0.08$^{***}$\\
&&(0.03)&(0.03)&(0.02)\\
Factory worker (bricks/stones/mill) &&-0.00&0.17$^{***}$&0.07$^{***}$\\
&&(0.02)&(0.03)&(0.01)\\
Small business                      &&0.22$^{***}$&0.18$^{***}$&0.05$^{***}$\\
&&(0.02)&(0.03)&(0.01)\\
Teacher                             &&0.05&0.22$^{***}$&0.09$^{***}$\\
&&(0.04)&(0.05)&(0.03)\\
Daily labourer                      &&-0.05$^{*}$&0.16$^{***}$&0.08$^{***}$\\
&&(0.03)&(0.03)&(0.02)\\
Wood cutter                         &&-0.03&0.15$^{*}$&0.12$^{***}$\\
&&(0.06)&(0.07)&(0.04)\\
Animal skin business                &&0.36&0.62$^{*}$&0.52$^{***}$\\
&&(0.23)&(0.28)&(0.15)\\   
\midrule
Control other careers && Y & Y& Y\\
Control village fix effect && Y & Y& Y\\                        
\hline \hline
\end{tabular}
}
\end{center}
\scriptsize{\justify Table \ref{table9} summarizes all careers that have a significant impact ($p<0.001$) on the likelihood of a household being selected by LASSO de-bias method from appendix table \ref{apdxtab11}. The first column uses whether one is predefined leaders as response variable, the second column uses whether one joins the micro-finance program as response variable and the third column uses whether one is selected by lasso as response variable. Standard errors in parentheses * $p<0.1$, ** $p<0.01$, *** $p<0.001$.}
\end{table}

\begin{table}[htbp]\caption{Second Stage: predefined leaders' careers} 
\begin{center}
\scalebox{0.8}{\label{table10}
\begin{tabular}{l*{4}{c}} 
\hline \hline
     &&  Predefined  & \multicolumn{2}{c}{ Selected by LASSO }\\
     \cline{4-5}
     && leaders      &  Cross-validation & De-bias \\
\midrule
Small business                      &&0.22$^{***}$&0.18$^{***}$&0.05$^{***}$\\
&&(0.02)&(0.03)&(0.01)\\
Tailor Garment worker               &&0.08$^{**}$&0.13$^{***}$&0.03\\
&&(0.03)&(0.04)&(0.02)\\
Hotel worker                        &&0.32$^{***}$&0.24$^{**}$&-0.02\\
&&(0.07)&(0.09)&(0.05)\\
Poojari                             &&0.53$^{***}$&0.15&-0.05\\
&&(0.13)&(0.16)&(0.09)\\
Veterinary clinic                   &&0.87$^{**}$&-0.03&0.00\\
&&(0.32)&(0.39)&(0.21)\\
Barber/saloon                       &&0.41$^{***}$&-0.01&-0.02\\
&&(0.10)&(0.12)&(0.06)\\
\midrule
Control other careers && Y & Y& Y\\
Control village fix effect && Y & Y& Y\\                        
\hline \hline
\end{tabular}
}
\end{center}

\scriptsize{\justify Table \ref{table10} summarizes all careers that have a significant impact (($p<0.01$)) on the likelihood of a household being selected by BSS as being among the ``predefined leaders" from appendix table \ref{apdxtab11}. The first column uses whether one is predefined leaders as response variable, the second column uses whether one joins the micro-finance program as response variable and the third column uses whether one is selected by lasso as response variable. Standard errors in parentheses * $p<0.1$, ** $p<0.01$, *** $p<0.001$.}
\end{table}

\begin{table}[htbp]\caption{Participation Rate when Targeting Different Leaders}
\begin{center}
\scalebox{0.8}{\label{table11}
\begin{tabular}{lcccc} 
\hline\hline
     && \multirow{2}{*}{In data} &  Predefined   &  LASSO       \\
     &&          & Leaders   &    Leaders\\
\midrule

Participation Rate     &&   \multirow{2}{*}{16\%} & \multirow{2}{*}{20\%} &  \multirow{2}{*}{33\%}    \\
(non-leaders) \\
\hline\hline
\end{tabular}
}
\end{center}
\scriptsize{\justify Table \ref{table11} reports the participation rate of non-leaders when all targeted leaders decided to join. 
The true participation rate in data is 16\%. If all predefined leaders decide to join,  the participation rate will increase to 20\%. If all LASSO detected leaders decide to join,  the participation rate will increase to 33\%.}
\end{table}

\section{Conclusions}
\label{sec_conclusions}
In this paper, I propose a novel spatial autoregression model which allows for \emph{heterogeneous} endogenous effects. Specifically, each individual has an individual-specific endogenous effect on her neighbors. My approach is useful for modeling a network with leaders and followers.   

I propose a set of instruments as well as a two stage LASSO (2SLSS) method to estimate my model. The instruments are constructed as a function of the independent variables and an adjacency matrix. I use a LASSO type estimator to select the valid instruments in the first stage and the influential individuals in the second stage. I propose a bias correction for my two-stage estimator following \cite{vandeGeer2014}. I derive the asymptotic normality for my ``de-bias" two-stage LASSO estimator and conduct robust inference including confidence intervals. 

My model can be extended to allow for more flexible structures. To apply LASSO, I assume that the number of influential individuals is sparse. I propose heterogeneous endogenous effects model with cliques to incorporate locally influential individuals, where the sparsity assumption is only applied to globally influential individuals. My model can also be extended to situations where there are multiple networks. I propose the use of the sparse group LASSO in my 2SLSS process. I derive the convergence rate and prove the consistency of selection for the sparse group LASSO estimator.

I apply my method to study villagers' decisions to participate in micro-finance programs in rural areas of Indian. I show that leaders in those villages have significant influence over their neighbors' decision to join the micro-finance program, and I provide rankings for the different social and economic networks among villagers. Based on how effectively each network spreads the impact of influential individuals' decisions, my method shows that some networks such as ``visit go-come" and ``borrow money" are much more effective in influencing villagers' decisions than other networks such as ``temple company" and ``medical help". I further show that individuals from certain careers such as agricultural workers, Anganwadi teachers and small business owners are more likely to influence other villagers and the ``predefined leaders" selected by BSS are different than the LASSO detected influential individuals.

\end{document}